\documentclass[sigconf,nonacm]{acmart}

\renewcommand\footnotetextcopyrightpermission[1]{}

\usepackage{booktabs}
\usepackage{multirow}
\usepackage{colortbl}
\usepackage{pifont}
\usepackage{listings}
\usepackage{subcaption}
\usepackage{tikz}
\usepackage{tabularx}
\usepackage{array}
\usepackage{xcolor}
\usepackage{graphicx}
\usepackage{enumitem}
\usepackage{textcomp}
\usepackage{xurl}

\newcommand{\squishlist}{
  \begin{list}{$\bullet$}
    { \setlength{\itemsep}{0pt}      \setlength{\parsep}{3pt}
      \setlength{\topsep}{3pt}       \setlength{\partopsep}{0pt}
      \setlength{\leftmargin}{1.0em} \setlength{\labelwidth}{1em}
      \setlength{\labelsep}{0.5em} } }
\newcommand{\squishend}{
    \end{list}  }

\usetikzlibrary{arrows.meta,positioning,shapes.geometric,fit,decorations.pathreplacing}

\lstdefinelanguage{TLAplus}{
  morekeywords={MODULE,EXTENDS,CONSTANT,VARIABLE,ASSUME,Init,Next,Spec,
    INVARIANTS,SPECIFICATION,TRUE,FALSE,IF,THEN,ELSE,LET,IN,
    UNCHANGED,EXCEPT,SUBSET,UNION,BOOLEAN},
  sensitive=true,
  morecomment=[l]{\\*},
  morestring=[b]",
  literate={/\\}{{$\wedge$}}1 {\\/}{{$\vee$}}1
           {=>}{{$\Rightarrow$}}1 {->}{{$\rightarrow$}}1
           {<<}{{$\langle$}}1 {>>}{{$\rangle$}}1
           {\\in}{{$\in$}}1 {\\notin}{{$\notin$}}1
           {\\A}{{$\forall$}}1 {\\E}{{$\exists$}}1
           {\\subseteq}{{$\subseteq$}}1
}
\lstdefinestyle{compactir}{
  basicstyle=\ttfamily\scriptsize,
  numbers=left,
  numberstyle=\tiny,
  numbersep=4pt,
  frame=single,
  breaklines=true,
  columns=fullflexible,
  xleftmargin=1.2em,
  framexleftmargin=1.0em,
  aboveskip=0.4em,
  belowskip=0.2em
}

\definecolor{bothfail}{HTML}{D32F2F}
\definecolor{specfail}{HTML}{F57C00}
\definecolor{implfail}{HTML}{FFA726}
\definecolor{speconly}{HTML}{FFD54F}
\definecolor{cellpass}{HTML}{4CAF50}
\definecolor{undercon}{HTML}{90A4AE}
\definecolor{notappl}{HTML}{E0E0E0}
\definecolor{partialres}{HTML}{64B5F6}

\definecolor{codeblue}{RGB}{0,70,180}
\definecolor{codered}{RGB}{180,20,60}
\definecolor{codegreen}{RGB}{0,130,60}
\definecolor{codegray}{RGB}{120,120,120}
\definecolor{codecyan}{RGB}{70,130,140}

\lstdefinelanguage{AgentIR}{
  sensitive=true,
  alsoletter={._-+},
  morekeywords=[1]{ProtocolIR.transition,ResponsibilityIR.control},
  morekeywords=[2]{id,actor,kind,trigger,reads,writes,modality,source_refs,tla_action,
                   control_id,spec_clause,policy_clause,sdk_enforcement,owner_type,evidence_level},
  morekeywords=[3]{TRUE,NOT_SPECIFIED,absent,composition-orphan,behavioral+source},
  morestring=[b]",
}

\lstdefinestyle{paperir}{
  language=AgentIR,
  basicstyle=\ttfamily\footnotesize,
  numbers=left,
  numberstyle=\tiny\color{codegray},
  numbersep=6pt,
  frame=none,
  xleftmargin=1.6em,
  columns=fullflexible,
  keepspaces=true,
  breaklines=true,
  showstringspaces=false,
  keywordstyle=[1]\bfseries\color{codered},
  keywordstyle=[2]\color{codeblue},
  keywordstyle=[3]\color{codegreen},
  stringstyle=\color{codeblue},
  commentstyle=\itshape\color{codecyan},
  aboveskip=0.4em,
  belowskip=0.2em,
  captionpos=b
}

\newcommand{\BOTHFAIL}{\cellcolor{bothfail!30}\textbf{BOTH}}
\newcommand{\SPECFAIL}{\cellcolor{specfail!30}SPEC}

\newcommand{\CELLPASS}{\cellcolor{cellpass!30}\ding{51}}

\newcommand{\NOTAPPL}{\cellcolor{notappl!30}--}

\newcommand{\sys}{AgentThread}
\newcommand{\tlaplus}{TLA\textsuperscript{+}}

\begin{document}
\pagestyle{plain}

\title{Formal Security Analysis of Agent Protocol Composition}

\author{Shenghan Zheng}
\affiliation{%
  \institution{Dartmouth College}
  \country{USA}
}

\author{Qifan Zhang}
\affiliation{%
  \institution{Palo Alto Networks}
  \country{USA}
}

\author{Zheng Zhang}
\affiliation{%
  \institution{Meta}
  \country{USA}
}

\author{Haonan Li}
\affiliation{%
  \institution{UC Riverside}
  \country{USA}
}

\author{Christophe Hauser}
\affiliation{%
  \institution{Dartmouth College}
  \country{USA}
}

\begin{abstract}
AI agent protocols define how agents use tools, delegate work, and
coordinate across software systems, but their security
requirements remain incomplete and inconsistently enforced across
deployments. We present \sys{}, a source-linked framework for security
assurance analysis of agent protocols,
from specification text to running SDKs. \sys{} contributes
a layered security scope, protocol-derived checks formalized as
TLA\textsuperscript{+} invariants, and a
two-phase checker that compiles protocol specifications into
model-checkable models and replays executable counterexamples
against real SDKs through protocol adapters. For each finding, \sys{} records the source text behind the
check and separates violated protocol requirements from missing
recommendations, hardening gaps, and unassigned cross-protocol
responsibilities.

Across five emerging agent protocols,
\sys{} identifies 35 specification-level findings, supports
them with 80 implementation tests against production SDKs and
reference servers, and finds 30 additional failures that emerge
only under protocol composition. We further show that only one
protocol enforces a security-relevant control in practice and no
protocol assigns enforcement for cross-protocol behavior.
Insecurity in agent protocols is therefore not only a
specification or implementation problem, but also a responsibility
gap across protocols, SDKs, and deployments.
\end{abstract}


\maketitle
\thispagestyle{plain}
\pagestyle{plain}

\section{Introduction}\label{sec:intro}

AI agent protocols are becoming software infrastructure.
MCP~\cite{mcp2024}, for example, defines how agents discover
tools and connect model-driven workflows to local files, cloud
services, peer agents, and developer environments. These protocols
are no longer narrow APIs around a single service: they mediate
tool access, delegation, and shared context across autonomous
workflows.

As these protocols mediate tool access, delegation, and shared context,
they create a software-engineering assurance
problem, and recent vulnerabilities show that these risks are no
longer hypothetical. MCP servers have already had command
injection, server-side request forgery, and arbitrary file read
bugs \cite{cve202549596, cve202568143, cve202568144}. Such single-server failures
matter, but they are not enough to
characterize current agent deployments. Agent runtimes routinely
compose multiple servers, so one server may supply untrusted
content while another provides file access or network egress. A
Snyk Labs study~\cite{snyk2025mcp} illustrates this composed failure: an agent
fetches an attacker-controlled webpage through one MCP server; the
hidden instruction enters the model context; the agent then invokes
a second MCP server that reads a local file and uses the first
server again to exfiltrate the result. No single
MCP server owns the whole failure. The unsafe behavior emerges from
composition, implicit authority transfer, missing consent, and weak
audit visibility across the agent runtime.

Existing approaches only partially address this problem. Classical
protocol verification is strong for message formats,
authentication, secrecy, and channel integrity
~\cite{lowe1996breaking,blanchet2001proverif,meier2013tamarin,basin2018formal,10.1145/3522582},
but agent deployments also require reasoning about semantic
payloads, delegated authority, and bridge/host behavior. Agent-security
work has documented prompt injection, tool abuse, and runtime
threats~\cite{greshake2023youve,zhan2024injecagent,debenedetti2024agentdojo,299563,DBLP:journals/csur/HeZYLZY26},
but typically studies attacks, benchmarks, or applications rather
than protocol specifications and SDK behavior. Recent
formal-methods work for agentic systems verifies plans or repairs
coordination artifacts~\cite{lee2025veriplan,xia2026tracefix};
our object of analysis is different: public communication
protocols, their implementations, and the software bridges that
compose them.


The central difficulty is maintaining an evidence chain from
protocol text to formal checks and implementation tests. Protocols
evolve quickly, their requirements are spread across prose, schemas,
SDK behavior, and security guidance, and many controls are delegated
to applications or developers. As a result, failures can arise from
three different places: a specification may omit or weaken a
security control, an SDK may not enforce the control that the
specification describes, or a composed deployment may create a
responsibility that no single protocol owns. A useful assurance method must ask what security properties agent
protocols should guarantee, how those guarantees can be checked
against specifications and implementations, and who is responsible
when they fail under composition.

We present \sys{}, a source-linked framework for security assurance and
composition analysis of agent protocols. \sys{} collects
protocol sources, extracts normative clauses into typed
intermediate records, compiles those records into \tlaplus{}
models~\cite{lamport2002specifying}, uses the TLC model
checker~\cite{yu1999model} to generate counterexample traces, and
replays those traces against real SDKs where an executable path
exists. The workflow is automated after IR acceptance: source
collection and protocol adapters remain human-supplied, while
extraction and IR construction are machine-assisted and
validation-gated. \tlaplus{} specifies state-machine behavior and invariants,
matching protocol states, tool calls, and cross-server traces. The
same intermediate records also track
ownership and enforcement evidence, so a finding can be interpreted
as a standards non-conformance bug, a recommendation gap, a
hardening gap, or an unassigned composition responsibility.

We study five protocols spanning tool invocation, multi-agent
delegation, identity-oriented networking, coding-agent access, and
agent lifecycle management: Model Context Protocol
(MCP)~\cite{mcp2024}, Agent-to-Agent Protocol
(A2A)~\cite{a2a2024}, Agent Network Protocol
(ANP)~\cite{anp2025}, Agent Communication Protocol
(ACP)~\cite{acp2025}, and Agent Client Protocol
(ACP-Client)~\cite{acpclient2025}. Across these protocols, \sys{}
identifies 35 specification-level findings, supports them with 80
implementation tests against production SDKs and MCP reference
servers, and finds 30 additional failures that appear only in
composed deployments. The results show that agent-protocol
insecurity is not only a matter of individual implementation bugs.
It is also a problem of fast-moving specifications, SDK enforcement
gaps, and responsibility left outside any one protocol boundary.

\emph{Contributions.}
This paper makes three contributions:
\begin{itemize}
    \item \textbf{A systematic formal analysis of agent
    protocols.} We define a layered security scope for agent
    communication protocols and derive checkable \tlaplus{}
    invariants for semantic content flow, delegated authority,
    session and capability state, audit, and composition.

    \item \textbf{A source-linked workflow from protocol text to tests.}
    \sys{} drafts clause, transition, property, and responsibility
    records; validates their source links and types; compiles
    accepted records to formal models; and turns executable TLC
    traces into SDK replay tests. The workflow preserves source
    references so findings can be reproduced and revised as
    protocols change.

    \item \textbf{An empirical study across protocols,
    implementations, and compositions.} We evaluate five protocols,
    production SDKs, MCP reference servers, and eight composed
    models, showing that many failures appear only when protocol
    behavior is checked against SDKs or composed through runtime
    bridges.
\end{itemize}

\section{Motivation}\label{sec:motivation}

\begin{figure}[t]
\centering
\includegraphics[width=\columnwidth]{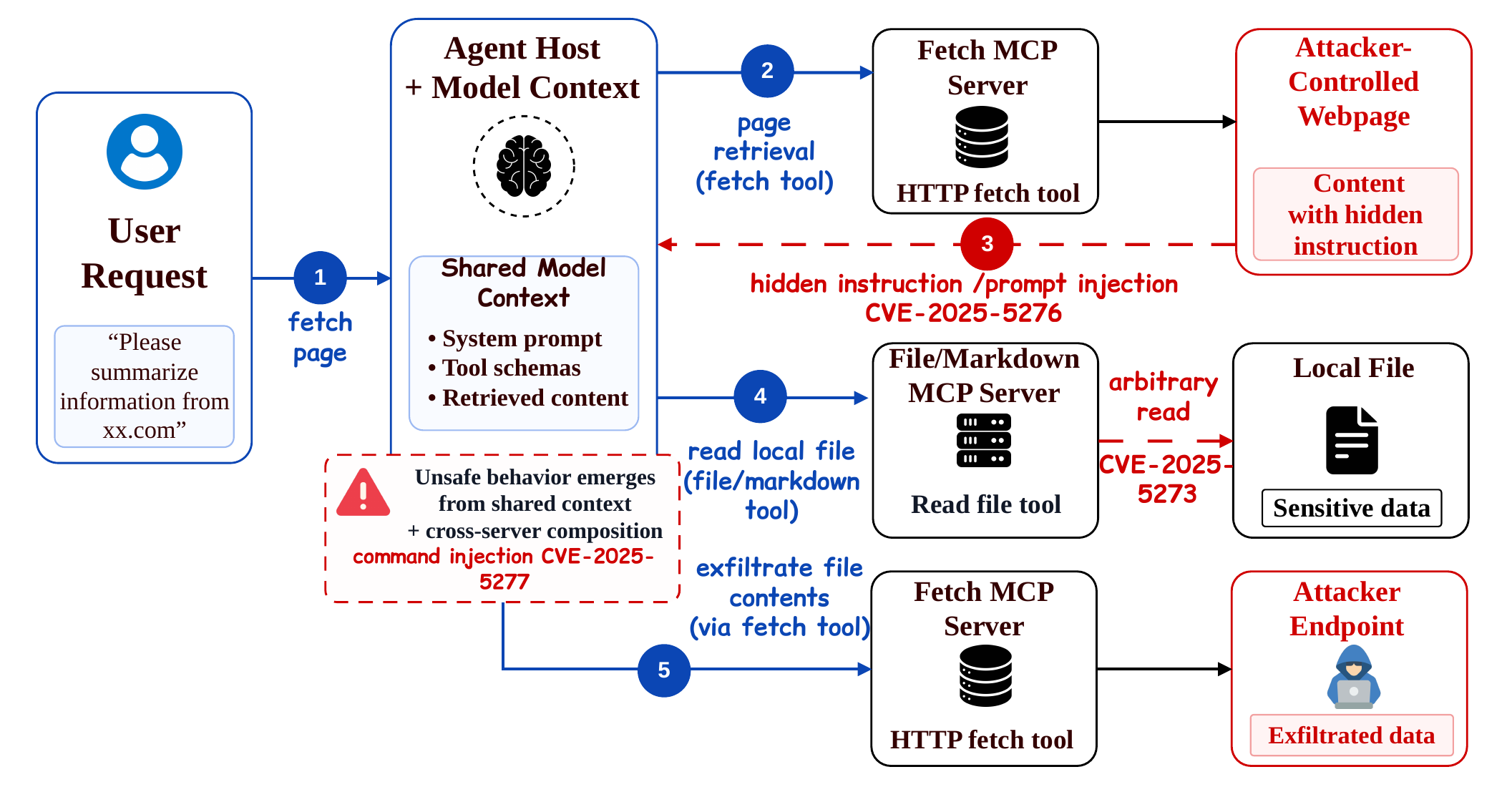}
\caption{Motivating multi-MCP server composition. A hidden instruction
fetched from the network can enter shared model context, steer a later
local-file read, and send the result through a network-capable server.}
\label{fig:mcp-chain}
\vspace{-0.9em}
\end{figure}

We use a real MCP multi-server failure pattern reported by Snyk
Labs~\cite{snyk2025mcp} as a running example and
Figure~\ref{fig:mcp-chain} shows the attack chain. An agent runtime/host
is connected to two MCP servers. One server fetches web content;
another converts local or remote files into Markdown. The vulnerable
point is not a malformed MCP message, but an unguarded bridge from
untrusted content to later tool authority: an attacker-controlled
page places hidden instructions into the agent's model context,
after which the agent uses the file-capable server to read local
data and the network-capable server to send the result to an
attacker-controlled endpoint.

The attack is possible because four controls fail together. First,
semantic content flows from an untrusted webpage into model state
and can steer later actions. Second, authority composes across protocol
servers: the same host can invoke a network-capable tool and a
file-capable tool in one workflow. Third, the consent boundary is
missing or too coarse, so the user's request to summarize a page is
not separated from later local-file access and outbound
transmission. Fourth, audit visibility is weak: each action appears
tool-valid in isolation, while the harmful intent is visible only in
the composed trace. These failures are increasingly relevant as
software-engineering agents become multi-turn, tool-using, and
multi-agent systems~\cite{gao2025more,li2025swe}.

\subsection{Text Becomes Tool Action}

The first step in the example is ordinary web retrieval. The
security-relevant event happens later, when hidden webpage text is
placed into the model context and interpreted as an instruction.
This is the core lesson of indirect prompt injection: untrusted
content can become operational intent once it reaches a
tool-integrated agent~\cite{greshake2023youve,zhan2024injecagent,299563}.
Agent benchmarks such as AgentDojo make the same point in a
controlled setting: defenses must reason about both the data an
agent observes and the tools it may invoke~\cite{debenedetti2024agentdojo}.

In the MCP chain, no message is malformed. Fetching a page,
reading a file, and sending a URL are each normal tool actions. The
bug appears because the runtime fails to bind later tool calls to
the user's original intent. A protocol-level analysis therefore
cannot stop at syntax or authentication. It must also ask whether
semantic content can cross into authority-bearing actions.

\subsection{Local Hardening Is Not Enough}

Snyk's practical guidance is correct: MCP servers should be treated
as untrusted third-party code, hardened and tested for familiar
bugs such as command injection, server-side request forgery, and
path traversal, and isolated from the main environment
~\cite{snyk2025mcp}. These practices reduce the blast radius of a
single server. They do not, by themselves, explain what happens
when two correctly functioning servers are composed by the same
agent runtime.

In Figure~\ref{fig:mcp-chain}, checking either server alone misses
the end-to-end path from attacker-controlled content to local-file
access and network exfiltration. The relevant state is distributed
across the host, the two clients, two servers, and the model context
that connects them. This motivates explicit composition models:
when protocol endpoints are joined by a runtime or bridge, security
properties must hold across the bridge, not only inside each
endpoint. This view matches recent observations that A2A--MCP
integration is moving from ad hoc glue code toward protocol-level
infrastructure~\cite{li2025glue}.

\subsection{The Owner Is Between Components}

The final difficulty is ownership. In the example, the fetch server
owns network access, the Markdown server owns file conversion, the
host owns tool selection, and the user interface may own consent.
No component, viewed alone, owns the rule: ``content fetched from
the network must not authorize a later local file read and outbound
send.'' A server maintainer can fix a path-validation bug, but the
composed control remains: who prevents the flow, who logs it, and
which specification requires it?

This is also a practical triage problem. Security programs expect
reproducible proof of concepts, demonstrable impact, and actionable
remediation paths~\cite{openai2026safetybugbounty}, but
agent-composition failures are multi-step, cross-component, and
stateful, so each step may look intended in isolation. The disputed
MCP STDIO case illustrates this gap: reports describe behavior treated
upstream as by design, while downstream systems still needed
mitigations once concrete exploit paths appeared
~\cite{ox2026mcparchitecturalflaw,csa2026mcpdesignrce}. Protocol
churn further makes one-time audits stale as prose, schemas, SDK
behavior, and bridge code change. We therefore need a workflow that
preserves source references, emits formal checks, regenerates replay
tests, and records whether any specification or SDK owns the relevant
cross-component control.

\section{Overview}\label{sec:overview}

\sys{} analyzes agent protocols as composed software systems rather than
as isolated message formats. The input is a set of authoritative
protocol artifacts: prose specifications, role-specific
client/server/agent documentation, schemas, examples, SDKs, and
reference implementations. The output is a set of source-linked
findings that explain (1) which source clause or missing clause
motivates a check, (2) whether the formal model admits a violating
execution, (3) whether an SDK or reference server exposes the same
behavior, and (4) who, if anyone, owns the relevant control.

\textbf{Threat model.}\label{sec:threat}
We assume a protocol-level Dolev--Yao adversary~\cite{dolev1983security}
augmented with agent-specific capabilities. The adversary can inject
instructions through tool outputs, webpages, or peer-agent messages;
forge or inflate capability metadata; and amplify authority through
delegation or tool chains. This captures a broader class of
agent-mediated security failures: untrusted external content,
capability metadata, or delegated context enters model state and
influences later API or tool calls, a pattern observed in real
LLM-integrated applications and systematized in agent benchmarks
~\cite{greshake2023youve,299563,zhan2024injecagent,debenedetti2024agentdojo}.

\begin{figure}[t]
\centering
\scriptsize
\begin{tikzpicture}[
  layer/.style={
    draw=#1!70,
    fill=#1!10,
    rounded corners=2pt,
    minimum width=0.70\columnwidth,
    minimum height=0.42cm,
    align=left,
    inner xsep=5pt,
    font=\scriptsize
  },
  group/.style={font=\scriptsize\bfseries, align=left},
  brace/.style={decorate, decoration={brace, amplitude=3pt}, thick, draw=#1!80}
]
\node[layer=blue]   (l5) at (0,0)        {\textbf{L5} Audit / accountability};
\node[layer=blue]   (l4) at (0,-0.48)    {\textbf{L4} Semantic operation controls};
\node[layer=green]  (l3) at (0,-1.10)    {\textbf{L3} Identity / capability binding};
\node[layer=green]  (l2) at (0,-1.58)    {\textbf{L2} Session lifecycle};
\node[layer=green]  (l1) at (0,-2.06)    {\textbf{L1} Message / wire-format integrity};

\draw[brace=blue]   (3.0,0.21) -- (3.0,-0.69);
\draw[brace=green]  (3.0,-0.89) -- (3.0,-2.27);

\node[group, text=blue!70!black, anchor=west]  at (3.12,-0.24) {Agent layer};
\node[group, text=green!60!black, anchor=west] at (3.12,-1.58) {RPC layer};
\end{tikzpicture}
\caption{Layered security scope used by \sys{}.}
\label{fig:security-scope}
\vspace{-0.8em}
\end{figure}
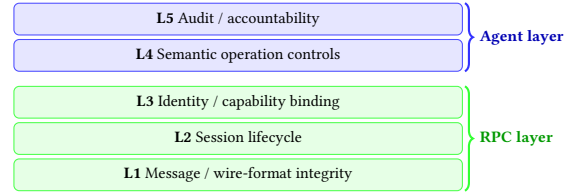

\textbf{Security scope.}
We treat transport security as an assumed substrate and focus on
the security obligations visible at the protocol and agent layers.
Figure~\ref{fig:security-scope} groups the scope into two parts.
The \emph{RPC layer} captures concerns inherited from
remote-procedure-call protocols: message and wire-format validity,
request/session correlation, and identity or capability binding
~\cite{rfc5531,rfc7861,rfc3552}. The \emph{agent layer} captures
the concerns exposed by Figure~\ref{fig:mcp-chain}: semantic
operation control and audit/accountability. In the motivating
example, the hidden webpage instruction steering a later file read
is a semantic-operation failure, while the cross-server
file-read-and-egress path is an audit/accountability failure because
composition controls are missing.
In composition, \sys{} lifts the same layers across bridge
boundaries: L1 checks well-formed translation, L2 checks lifecycle
cascade, L3 checks authority monotonicity, L4 checks
semantic-to-authority isolation, and L5 checks audit/provenance
continuity.

\textbf{Design rationale.}
The workflow is organized around three frictions exposed by the
motivating example. First, protocol requirements are written across
natural language, role-specific documentation, schemas, examples,
and SDK behavior, so \sys{}
separates source collection, clause extraction, typed IR construction,
and \tlaplus{} compilation while preserving source provenance.
Second, responsibility is distributed across specifications, SDKs,
hosts, users, and bridges, so the IR records not only protocol state
but also ownership and enforcement evidence. Third, formal
counterexamples are useful to practitioners only when they can be
reproduced, so \sys{} lowers TLC traces through protocol adapters
into implementation replays or source/type checks, keeping the
formal trace as the test oracle.

\begin{figure}[t]
\centering
\includegraphics[width=\columnwidth]{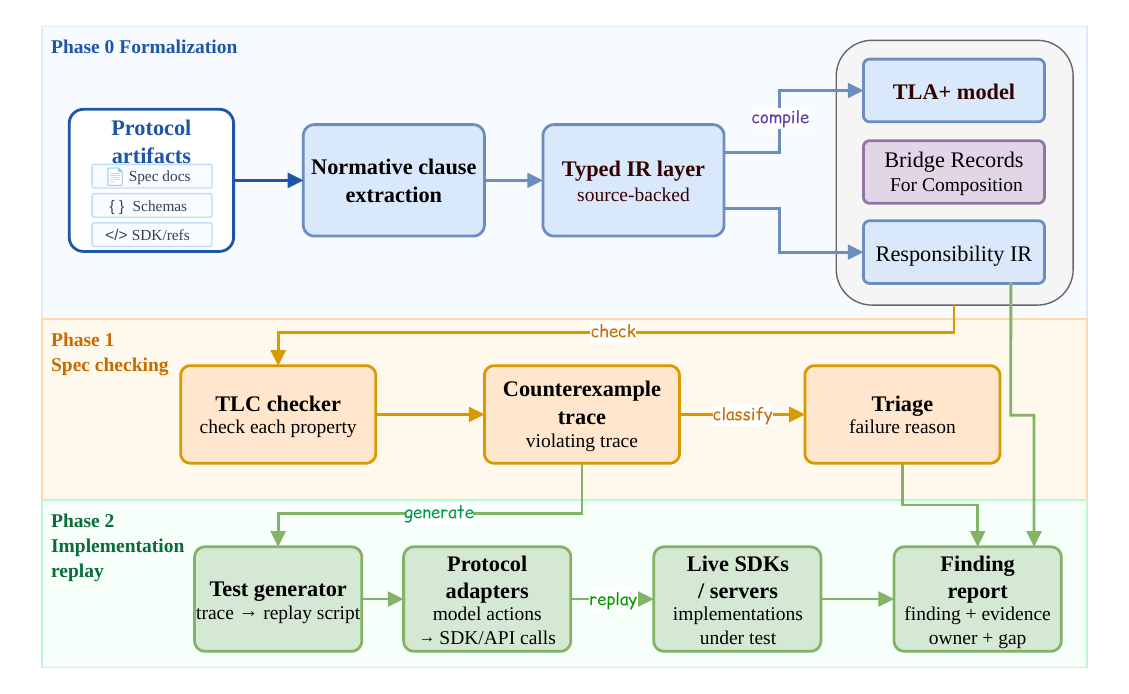}
\caption{\sys{} workflow. Protocol artifacts are normalized into
source-backed clauses and typed IR, compiled into formal checks, and
replayed against implementations when counterexamples are executable.}
\label{fig:workflow-pipeline}
\vspace{-0.9em}
\end{figure}

\textbf{Workflow.}
Figure~\ref{fig:workflow-pipeline} summarizes the end-to-end workflow.
\sys{} first normalizes collected protocol artifacts into
source-backed clauses and typed IR records. It then compiles the IR with bridge records
into \tlaplus{} models and runs TLC to produce either verified
checks or counterexample traces. When a counterexample corresponds
to executable behavior, \sys{} lowers the trace through a protocol
adapter into an SDK or reference-server replay; otherwise, it
records source/type-level evidence. The final report links each
finding back to its source text, formal trace, implementation
evidence, and responsible owner.

\section{The \sys{} Framework}\label{sec:framework}

This section describes how \sys{} turns the scope in
\S\ref{sec:overview} into executable checks. We first map the RPC
and agent-layer scope to verification targets
(\S\ref{sec:framework:aasm}), then describe the Protocol IR and
two-phase assurance checker
(\S\ref{sec:framework:ir}--\S\ref{sec:framework:replay}). We then
introduce the Responsibility IR
(\S\ref{sec:framework:responsibility-ir}) and the composition
methodology (\S\ref{sec:framework:composition}).

\subsection{Layer-Derived Checks}\label{sec:framework:aasm}

Starting from the five layers in \S\ref{sec:overview}, \sys{} asks
whether the protocol text provides enough normative structure to
state a safety property and whether the SDK exposes enough behavior
to test it. We instantiate the layers as a compact set of
verification targets. RPC-layer targets include wire-format
integrity, session lifecycle, identity verifiability, capability
attestation, credential or registry integrity, and fail-secure
defaults. Agent-layer targets include prompt/content integrity,
consent explicitness, audit completeness, delegation monotonicity,
and composition safety. The set is intentionally small: it is meant
to cover the recurring security state needed by current agent
protocols, not to define a universal checklist. Each target becomes
a protocol-specific \tlaplus{} invariant over the state variables
and actions generated from the Protocol IR.

\textbf{Property taxonomy.}\label{sec:framework:taxonomy}
To avoid calling every failed check a protocol violation, each
property carries a source-evidence label. \textbf{Spec-Mandated}
properties are backed by MUST-level protocol text; a failed check is
standards non-conformance. \textbf{Spec-Recommended} properties are
backed by SHOULD-level or best-practice text; a failed check is a
recommendation gap. \textbf{Framework hardening} properties are
required by our layer-derived checks but absent from the documentation;
a failed check is a design gap. \textbf{Layer completeness}
properties capture missing architectural obligations at a layer.
This taxonomy is used throughout the evaluation: only MUST-backed
failures are reported as protocol violations. Table~\ref{tab:taxonomy-examples}
shows representative classifications across the layers.

\begin{table}[t]
\centering
\caption{Examples of how layer-derived checks are classified.}
\label{tab:taxonomy-examples}

\footnotesize
\setlength{\tabcolsep}{2.5pt}
\renewcommand{\arraystretch}{1.12}

\begin{tabularx}{\columnwidth}{@{}
  >{\raggedright\arraybackslash}p{0.36\columnwidth}
  >{\centering\arraybackslash}p{0.13\columnwidth}
  >{\centering\arraybackslash}p{0.08\columnwidth}
  >{\raggedright\arraybackslash}X
@{}}
\toprule
\textbf{Layer} & \textbf{Proto.} & \textbf{Cls.} & \textbf{Evidence} \\
\midrule
RPC-L1: Message / wire-format integrity
& ANP
& SM
& MUST encrypt messages \\

RPC-L2: Session lifecycle
& MCP
& SM
& MUST reject unauthorized requests \\

RPC-L3: Identity / capability binding
& ANP
& SM
& MUST verify DID identity \\

RPC-L3: Identity / capability binding
& MCP
& FH
& Capability attestation NS \\

Agent-L4: Semantic operation control
& MCP
& SR
& SHOULD obtain consent \\

Agent-L5: Audit / accountability
& MCP
& SR
& SHOULD log security events \\

\bottomrule
\end{tabularx}

\vspace{2pt}
\begin{minipage}{\columnwidth}
\footnotesize
\emph{Class types:} SM = Spec-Mandated; SR = Spec-Recommended;
FH = Framework hardening; LC = Layer completeness;
NS = Not Specified.
\end{minipage}
\end{table}

\textbf{Counterexample triage.}\label{sec:framework:triage}
TLC counterexamples are first triaged at the specification level as
\textbf{Spec-Fail} (the specification permits an unsafe execution),
\textbf{Model-Fail} (the abstraction introduced an unsupported
behavior), or \textbf{Ambiguity-Fail} (the source text does not
determine a unique behavior). Phase~2 then adds implementation
evidence: a replayed or source-confirmed implementation failure is
reported as \textbf{Impl-Fail} when the specification forbids the
behavior, and \textbf{Both-Fail} when both the specification and the
implementation admit it. This separation prevents TLC results from
being mistaken for implementation evidence and makes false positives
explicit instead of hiding them in the analysis workflow.

\subsection{Protocol IR: From Specifications to a Typed
Representation}\label{sec:framework:ir}

\sys{} formalizes protocol behavior through a Protocol IR
rather than translating informal prose directly into \tlaplus{}.
The IR serves two purposes. First, it makes the formalization
provenance-preserving: every \tlaplus{} action and
invariant is linked to a specific extracted clause. Second, it
makes the automation boundary explicit: source collection and
adapters are human-supplied; Steps~2--3 are machine-assisted and
validation-gated; and compilation, model checking, trace extraction,
and replay are automated after IR acceptance. We describe each step
below using the MCP multi-server
flow from Figure~\ref{fig:mcp-chain} as the running example.

\textbf{Step 1: Source collection (manual).}
For each protocol we collect authoritative artifacts: prose
specifications, role-specific(client/server/agent) documentation, schemas, examples, SDK
documentation, and reference implementations. Because obligations
are often split across these sources, we assign precedence
(normative role/spec text $>$ schema $>$ SDK/reference code $>$
example) and record disagreements unresolved with precedence as ambiguities rather
than silently choosing one interpretation.

\textbf{Step 2: Normative clause extraction (machine-assisted,
validation-gated).}
The extractor captures clauses that affect behavior, safety,
authority, lifecycle, or trust boundaries. Each clause is stored
with its modality (\textbf{must}, \textbf{should}, \textbf{may},
or \textbf{not specified}, per RFC~2119~\cite{rfc2119}), a source
reference (URL + anchor), an actor binding---that is, the
principal the clause applies to, a client, server, agent, bridge,
or adversary as distinct from the human users these principals
serve---and an explicit ambiguity flag. The
ambiguity flag is set to true when a clause admits more
than one legal reading; Step~4 will render ambiguous clauses as
nondeterministic \tlaplus{} actions so that TLC explores both
interpretations. In the MCP running example, clauses governing tool
results, user consent, and server responsibility become distinct IR
clause records with source references and modality tags; the
content-integrity property then records whether the relevant
obligation is mandated, recommended, or absent. Static rules split
sections and identify RFC-style modalities; machine assistance handles
semantic cases such as actor/topic tagging and compound clauses.
Records are rejected if their source span is absent from the
document or their modality/enums are ill-formed.
The resulting records include the source links needed to inspect the
corresponding IR chain and compiled \tlaplus{} action.

\textbf{Step 3: Typed IR construction (machine-assisted,
validation-gated).}
Unlike prior staged extraction pipelines such as Pacheco et
al.~\cite{pacheco2022}, Hermes~\cite{hermes2024}, and
PROSPER~\cite{prosper2023}, which primarily recover protocol state
machines or logical transition formulas from specification text, we
use the IR not only to encode protocol behavior but also to preserve
source modality, layer-derived security properties, and
responsibility or enforcement ownership. The Protocol IR has three record
families: clause records preserve provenance, transition records
encode protocol behavior---who acts, under what conditions, with
what state update and modality---and property records attach the
taxonomy labels from \S\ref{sec:framework:taxonomy}. This lets downstream
findings be interpreted directly as standards non-conformance,
hardening gaps, or under-specification without
re-inspecting the source text. These records are typed and
structured (one record per clause, transition, or property).
Human mapping proposes transition components, while
templates instantiate common layer-derived property records.
Validators check source references, action kinds, property
classes, and required \tlaplus{} action/invariant names before
draft IR artifacts are emitted.

\textbf{Step 4: IR-to-\tlaplus{} compilation (automated).}
A syntax-directed compiler turns the IR into a \tlaplus{}
specification. Actors become quantified principals, triggers and
preconditions become guarded actions, state writes become primed
variable updates, and modality determines whether a property becomes
a safety invariant, hardening check, or under-specification finding.
Clauses flagged as ambiguous produce nondeterministic branches so
that TLC explores all interpretations. The compiler also preserves
the action kind: protocol actions are transitions licensed by the
specification, environment actions model nondeterministic but
non-malicious events (e.g., token expiry), and adversary actions
model explicit attacker capabilities from \S\ref{sec:threat}.
Table~\ref{tab:ir-tla-mapping} summarizes the compilation mapping.

\begin{table}[t]
\centering
\caption{IR-to-\tlaplus{} compilation mapping.}
\label{tab:ir-tla-mapping}
\footnotesize
\setlength{\tabcolsep}{4pt}
\begin{tabularx}{\columnwidth}{@{}l>{\raggedright\arraybackslash}X@{}}
\toprule
\textbf{IR construct} & \textbf{\tlaplus{} construct} \\
\midrule
\texttt{actor} & Quantified principal in $\forall$ / $\exists$ \\
\texttt{trigger} & Guarded-action enabled predicate \\
\texttt{preconditions} & Action enabled conjunct \\
\texttt{state\_writes} & Primed variable update \\
\texttt{kind: Protocol} & Action in \texttt{ProtocolNext} \\
\texttt{kind: Environment} & Action in \texttt{EnvNext} \\
\texttt{kind: Adversary} & Action in \texttt{AdvNext} \\
\texttt{MUST} & Safety invariant \\
\texttt{SHOULD} & Annotated hardening invariant \\
\texttt{NOT\_SPECIFIED} & Under-specification finding \\
\texttt{ambiguity: true} & Nondeterministic action branches \\
\bottomrule
\end{tabularx}
\end{table}

\subsection{Responsibility IR: From Specifications to Responsibility
Records}\label{sec:framework:responsibility-ir}

In addition to the Protocol IR, \sys{} extracts Responsibility IR
records that track who owns each control and whether the SDK
enforces it. Each record links the governing clause, modality,
security-policy clause, claimed principal, observed SDK enforcement,
ownership/gap type, and evidence. Listing~\ref{lst:ir-records}
shows a compact MCP example: the transition record is compiled into
\tlaplus{}, while the responsibility record drives ownership and
enforcement analysis.

\begin{lstlisting}[style=paperir,
caption={Example Protocol and Responsibility IR records.},
label={lst:ir-records}]
ProtocolIR.transition {
  id: "MCP-L4-read-after-content",
  actor: "agent_host", kind: "Protocol",
  trigger: "invoke file tool",
  reads: ["model_context", "tool_capabilities"],
  writes: ["local_file_read := TRUE"],
  modality: "NOT_SPECIFIED",
  source_refs: ["MCP-C-tool-results"],
  tla_action: "ReadLocalFile"
}
ResponsibilityIR.control {
  control_id: "MCP-L5-composition-accountability",
  spec_clause: "MCP-C-tool-results",
  modality: "NOT_SPECIFIED",
  policy_clause: "content cannot authorize file reads or egress",
  sdk_enforcement: "absent",
  owner_type: "composition-orphan",
  evidence_level: "behavioral+source"
}
\end{lstlisting}

Responsibility records are generated from the same source-backed
clauses as the Protocol IR and then validated against SDK or
reference-server evidence. For each security target, \sys{} records
the clause and modality, any corresponding security-policy text, the
principal named by the specification, the observed enforcement
mechanism, and the resulting gap type: ownership, split-duty,
enforcement, scope, or composition-orphan (see
\S\ref{sec:eval:responsibility} for empirical use). In spirit this
extends NIST OSCAL~\cite{oscal} from fixed control catalogs to
protocol-derived normative clauses, but the records are specialized
for agent protocols whose controls often span specifications, SDKs,
hosts, and bridges.

\subsection{Two-Phase Assurance Checker}\label{sec:framework:replay}

\textbf{Phase~1: Spec-level analysis.}
Phase~1 checks each layer-derived invariant independently using TLC.
Rather than bundling all invariants into a single conjunction
(which would cause TLC to stop at the first violation), we generate a
separate TLC configuration for each invariant, producing one
counterexample per violated check. The spec-level analysis
parses TLC output to extract (1)~which invariant was violated,
(2)~the depth of the counterexample, and (3)~the full state trace
with variable values at each step. Each counterexample is
classified using the triage taxonomy
(\S\ref{sec:framework:triage}), which treats \textbf{Model-Fail}
and \textbf{Ambiguity-Fail} findings as first-class outcomes.

\textbf{Phase~2: Implementation-level testing.}
Phase~2 converts TLC counterexample traces into executable test
cases when the trace has an executable implementation path. The
same Protocol IR property that generated the invariant in Phase~1
also provides the Phase~2 oracle, giving one chain from a source
clause to a replayed SDK test. Protocol adapters map \tlaplus{}
actions to concrete SDK or API calls; composed traces use a harness
that hosts both protocol stacks plus a bridge module and dispatches
each step to the relevant adapter. When a trace cannot be replayed
directly, \sys{} records source/type-level evidence for the absent
mechanism instead. This counterexample-to-test workflow
follows the tradition of rigorous protocol conformance
testing~\cite{bishop2005rigorous}, but uses formally checked
\tlaplus{} traces as the oracle source.


\subsection{Composition Methodology}\label{sec:framework:composition}

A key strength of the Protocol IR is that it composes:
when two protocols meet through a bridge, conductor, or agent
runtime, we can build a composed \tlaplus{} model by reusing the
per-protocol IR transitions and adding bridge-level actions that
transfer state across protocol boundaries. Each bridge action is
paired with a finite capability contract that states which context,
credential, tool, consent, delegation, or audit rights the
bridge/runtime may read, write, or exercise. This is the same
workflow as \S\ref{sec:framework:ir}--\S\ref{sec:framework:replay};
composition does not require a separate framework, only
additional IR records. We restrict analysis to pairwise
combinations; triples would multiply TLC's state space and obscure
which bridge action created a failure. Pairwise models keep the
composition boundary explicit.

\textbf{Composed-model construction.}
A composed \tlaplus{} model for protocols $P_1$ and $P_2$ is built
from four ingredients:
(1)~the IR transitions of $P_1$, imported as-is and namespaced;
(2)~the IR transitions of $P_2$, imported as-is and namespaced;
(3)~a set of bridge actions that model how the
conductor, relay, or agent runtime carries state from $P_1$ to
$P_2$ (e.g., ``the MCP tool output becomes A2A task input'');
(4)~a set of composition safety invariants that assert
that security properties of $P_1$ or $P_2$ are preserved under
the bridge actions. Bridge actions are declared as
\texttt{kind=Protocol} when they are licensed by either
specification, \texttt{kind=Environment} when they are
deployment-level glue that neither specification normatively
defines, and \texttt{kind=Adversary} when they model a
bridge-level compromise (ADV-1 through ADV-3 acting through the
bridge).
In the generated \tlaplus{} model, the bridge contract is a finite
capability set used as an action guard. For example, an MCP tool
result can be forwarded into agent context only if the bridge has
the corresponding context-write right, and a file-read-plus-egress
trace requires both file and network rights. Counterexamples
therefore identify not only that state crossed a protocol boundary,
but also which bridge/runtime capability made the transfer possible.

\textbf{Invariant design.}
For each protocol pair, \sys{} instantiates security obligations
from the layer-lifted templates introduced in
\S\ref{sec:overview}. The templates lift each local layer obligation
across the bridge boundary: L1 checks translation well-formedness,
L2 lifecycle propagation, L3 authority monotonicity, L4
semantic-to-authority isolation, and L5 audit/provenance continuity.
We distinguish two obligation classes. \emph{Boundary-risk}
obligations ask whether composition exposes a cross-protocol attack
path, such as content-to-authority flow or stale delegated
authority. \emph{Bridge-contract preservation} obligations are
positive controls: they check that a bridge does not mint
authentication, rewrite grant baselines, hide delegation records, or
produce local side effects without source state allowed by its
contract. Model-specific names such as \emph{InjectionToFileWrite}
are concrete instantiations of these templates.

We check model-validation invariants separately. These cover
state-domain separation, lifecycle/counter bounds, authority
domains, bridge-source guards, and audit-counter bounds. They
validate that the composed model is well-scoped and are excluded
from the security-obligation denominator in
\S\ref{sec:eval:composition}.

\section{Implementation}\label{sec:implementation}

\textbf{Adapters and replay loop.}
Each executable target has a small adapter module that maps the
\tlaplus{} action names emitted by Phase~1 to concrete SDK or
reference-server calls. Adapters are hand-written because each
target exposes a different call surface, but they share a common
base class that implements counterexample replay: given a trace,
the base walks the action sequence, dispatches each action to the
adapter, and after each step evaluates the invariant predicate that the
generating Protocol IR property points to. The composition
harness defined in \S\ref{sec:framework:replay} reuses this
machinery, hosting two adapters plus a bridge module in one process.
The bridge module implements the same capability contract used in
the generated \tlaplus{} model, so replayed steps identify both the
crossed boundary and the bridge/runtime right that enabled it.

\textbf{Automation boundary.}
Source collection and adapter implementation are manual setup
steps. The extraction front end is machine-assisted: it proposes
source-backed clauses, transition skeletons, property records,
responsibility records, and bridge capability records; rejects
malformed outputs through validators; and emits draft
\texttt{ir\_clauses.json}, \texttt{ir\_transitions.json}, and
\texttt{ir\_properties.json}. Accepted IR is the boundary for
downstream automation. From accepted IR onward, model checking,
output parsing, counterexample-trace extraction, and replay against
existing adapters are automated.

\textbf{Implementation footprint.}
The implementation consists of source-extraction scripts,
validators, Protocol and Responsibility IR files, bridge capability
records, generated \tlaplus{} specifications, a TLC driver, adapters,
and replay or source-check tests. The evaluation reports the corpus
size, checked model bounds, and replay coverage used in this study.

\section{Evaluation}\label{sec:eval}

We evaluate \sys{} through four research questions:

  \textbf{RQ1 (Specification)} What specification-level security gaps appear in
  individual agent-protocol specifications?
  
  \textbf{RQ2 (Implementation)} Which model-checked findings reproduce against real
  SDKs or reference servers?
  
  \textbf{RQ3 (Composition)} What failures appear only when protocols are composed
  through an agent runtime, relay, or bridge?
  
  \textbf{RQ4 (Responsibility)} Who owns and enforces the controls involved in these
  findings?

\subsection{Evaluation Setup}\label{sec:eval:setup}

Every \tlaplus{} action and invariant is linked to IR records. IR
counts (clauses / transitions / properties) are
MCP~37/11/7, A2A~17/6/6, ANP~16/11/7, ACP-Cap~8/7/6, and
ACP-Client~15/9/7. Each property carries a taxonomy tag
(\textbf{spec-mandated}, \textbf{spec-recommended},
\textbf{framework hardening}, or \textbf{layer completeness}).
The TLC driver is
v2.19; the SDK versions are MCP~v1.26.0, A2A~v0.3.25,
ACP-Cap~v1.0.3, ANP~v0.7.2, and ACP-Client~v0.9.0. Additionally,
we test three official MCP reference servers
(mcp-server-fetch, mcp-server-sqlite, mcp-server-git) as
real-world proof-of-concept targets.

Table~\ref{tab:coverage} summarizes the evidence available for each
protocol. We use these evidence levels to scope claims: behavioral
replay and source inspection are reported separately rather than
treated as interchangeable implementation evidence.

\begin{table}[t]
\centering
\caption{Evaluation coverage by protocol and analysis axis.}
\label{tab:coverage}
\small
\setlength{\tabcolsep}{3.5pt}
\resizebox{\columnwidth}{!}{%
\begin{tabular}{@{}lccccc@{}}
\toprule
\textbf{Axis} & \textbf{MCP} & \textbf{A2A} & \textbf{ANP} & \textbf{ACP-Cap} & \textbf{ACP-Client} \\
\midrule
Protocol IR         & \ding{51} (37) & \ding{51} (17) & \ding{51} (16) & \ding{51} (12)  & \ding{51} (15) \\
TLA$^{+}$ model     & \ding{51}      & \ding{51}      & \ding{51}      & \ding{51}    & \ding{51}      \\
TLC model checking  & \ding{51}      & \ding{51}      & \ding{51}      & \ding{51}    & \ding{51}      \\
SDK analysis/tests      & \ding{51} (18) & \ding{51} (13) & \ding{51} (8) & \ding{51} (9)& \ding{51} (16)   \\
\shortstack[l]{Server\\proof-of-concept} & \ding{51} (7)  & --             & --             & --           & --             \\
Composition models  & \ding{51} (5)  & \ding{51} (4)  & \ding{51} (2)  & \ding{51} (2)& \ding{51} (1)  \\
Composition behavioral & \ding{51} (1) & --           & --             & --           & \ding{51} (1)  \\
Responsibility IR   & \ding{51} (14) & \ding{51} (9)  & \ding{51} (5)  & --           & \ding{51} (7)  \\
\midrule
\textbf{Technical evidence} & L3   & L3   & L3   & L2   & L2 \\
\bottomrule
\end{tabular}
}

\vspace{2pt}
{\footnotesize Numbers indicate record/test
counts. L0~=~spec only, L1~=~source
inspection, L2~=~behavioral confirmation, L3~=~reproducible
exploit chain.}
\end{table}

\textbf{Role and bounds of TLC.}
Clause extraction can flag absent or weak requirements, but it does
not show whether the corresponding bad state is reachable or provide
a replayable trace. We therefore use TLC as a bounded
counterexample generator with an operation-count bound of 12:
each invariant is checked in a separate configuration, so
a failed check returns the shortest reachable trace and concrete
state values for Phase~2. The constants in
our TLC configurations are intentionally small but instantiate the
attack schemas we report: the required principals, sessions,
capabilities, messages, and, for composition, a bridge. Larger
constants add symmetric interleavings but are not needed to witness
these findings. This matches recent agent-protocol incidents and
the Snyk motivating example which requires one tool per server and Git-server CVEs turn on a single exposed endpoint or tool boundary rather than large deployments
~\cite{snyk2025mcp,cve202549596,cve202568143,cve202568144,cve202568145}.
Table~\ref{tab:tlc-performance} reports the explicit TLC bounds used
for each model. These bounds are model-specific because each model
is a verification slice for a particular family of invariants, not a
cartesian product of all protocol dimensions. For example, in the standalone MCP
model, sessions, tools, and messages remain explicit because the
single-protocol checks reason about those dimensions directly. But in some composed models, the MCP side is specialized to the minimal
role-level path needed for the cross-protocol invariant like one session or one tool-response path to make only the message
or operation bound tunable.

\begin{table}[t]
\centering
\caption{TLC model checking performance(operation-count bound = 12). Failed-check times are small
because TLC terminates at the first counterexample.}
\label{tab:tlc-performance}
\footnotesize
\setlength{\tabcolsep}{3pt}
\begin{tabular}{@{}l>{\raggedright\arraybackslash}p{0.36\columnwidth}ccc@{}}
\toprule
\textbf{Model} & \textbf{Constants} & \textbf{Viol.} & \textbf{Viol.\ Time} & \textbf{Sec.\ PASS} \\
\midrule
MCP         & S=3, T=3, M=8      & 7/9 & $<$3\,min each & \ding{51} \\
A2A         & A=4, C=3, M=8      & 4/6 & $<$5\,min each & \ding{51} \\
ANP         & A=3, M=6           & 4/7 & $<$2\,min each & \ding{51} \\
ACP-Cap     & A=3, C=3, M=6      & 4/6 & $<$1\,min each & \ding{51} \\
ACP-Client  & Ops=10             & 5/7 & $<$2\,min each & \ding{51} \\
\midrule
MCP+A2A     & S=2, A=3, C=3, M=8 & 7/9 & $<$5\,min each & \ding{51} \\
MCP+MCP     & M=8                & 5/7 & $<$2\,min each & \ding{51} \\
MCP+ACP-Cap & M=6                & 3/4 & $<$1\,min each & \ding{51} \\
A2A+ACP-Cap & A=3, C=3, M=6      & 3/5 & $<$4\,min each & \ding{51} \\
A2A+A2A     & A=3, C=3, M=6      & 2/4 & $<$5\,min each & \ding{51} \\
MCP+ACP-Cl  & Ops=10             & 4/6 & $<$1\,min each & \ding{51} \\
ANP+MCP     & Ops=8              & 3/4 & $<$1\,min each & \ding{51} \\
ANP+A2A     & Ops=8              & 3/4 & $<$3\,min each & \ding{51} \\
\bottomrule
\end{tabular}

\vspace{2pt}
\parbox{\columnwidth}{\footnotesize
S = sessions, T = tools, A = agents, C = capabilities, M = messages,
Ops = operations. Sec.\ PASS means at least one security obligation
passes; model-validation checks are separate and pass under the same bound.
}
\end{table}

\subsection{RQ1: Specification-Level Findings}\label{sec:eval:spec}

Table~\ref{tab:conformance-matrix} presents the complete $5 \times 11$
security-check matrix. Across 55~protocol--check cells, we observe
35~spec-level findings: 12 are confirmed by implementation tests,
23 are specification-only, 10 pass, and 10 are not applicable or not
checked because of protocol scope (e.g., MCP has no delegation model,
so the delegation check is N/A). The taxonomy introduced in \S\ref{sec:framework:taxonomy} determines how to
read each failed check: spec-mandated findings are protocol non-conformance,
spec-recommended findings are recommendation gaps, and framework
hardening or layer-completeness findings are design gaps.

\begin{table}[t]
\centering
\caption{Security-check matrix by layer-derived check. \BOTHFAIL{}~=~failed at
both specification/model and implementation levels; \SPECFAIL{}~=~
specification/model-level finding only; \CELLPASS{}~=~pass;
\NOTAPPL{}~=~not applicable.}
\label{tab:conformance-matrix}
\scriptsize
\setlength{\tabcolsep}{2.5pt}
\renewcommand{\arraystretch}{1.08}
\resizebox{\columnwidth}{!}{%
\begin{tabular}{@{}lccccc@{}}
\toprule
\textbf{Layer-derived check} & \textbf{MCP} & \textbf{A2A} & \textbf{ANP} & \textbf{ACP-Cap} & \textbf{ACP-Cl} \\
\midrule
RPC-L1 message / wire format      & \SPECFAIL & \CELLPASS & \CELLPASS & \CELLPASS & \CELLPASS \\
RPC-L2 session lifecycle          & \SPECFAIL & \CELLPASS & \CELLPASS & \CELLPASS & \SPECFAIL \\
RPC-L3 identity binding           & \SPECFAIL & \CELLPASS & \CELLPASS & \SPECFAIL & \NOTAPPL \\
RPC-L3 capability attestation     & \BOTHFAIL & \CELLPASS & \NOTAPPL & \SPECFAIL & \NOTAPPL \\
RPC-L3 credential / registry      & \BOTHFAIL & \BOTHFAIL & \SPECFAIL & \BOTHFAIL & \SPECFAIL \\
Agent-L4 delegation monotonicity  & \NOTAPPL  & \BOTHFAIL & \NOTAPPL & \SPECFAIL & \NOTAPPL \\
Agent-L4 content integrity        & \BOTHFAIL & \NOTAPPL  & \NOTAPPL & \SPECFAIL & \SPECFAIL \\
Agent-L4 consent explicitness     & \BOTHFAIL & \BOTHFAIL & \SPECFAIL & \BOTHFAIL & \SPECFAIL \\
Agent-L5 audit completeness       & \BOTHFAIL & \BOTHFAIL & \SPECFAIL & \BOTHFAIL & \SPECFAIL \\
Cross-layer fail-secure defaults  & \SPECFAIL & \SPECFAIL & \SPECFAIL & \SPECFAIL & \SPECFAIL \\
Agent-L5 composition safety       & \SPECFAIL & \SPECFAIL & \SPECFAIL & \NOTAPPL  & \NOTAPPL \\
\bottomrule
\end{tabular}
}

\smallskip
\raggedright\scriptsize
\end{table}

No protocol passes all checks. Two gaps appear across all five
protocols: audit completeness and credential/registry integrity are
either violated, only recommended, unenforced, or underspecified.
We therefore separate failures by source evidence. MCP's
content-integrity check is \textbf{spec-mandated}: the specification
requires sanitization, but the SDK does not comply. A2A's delegation,
consent, audit, and credential-lifecycle checks and the ACP-Client
findings are instead \textbf{framework hardening} gaps because the
specifications leave those requirements absent or permissive. Only
MUST-backed failures are protocol non-conformance; the rest are
recommendation, hardening, or design gaps.

The most severe single-protocol result is ACP-Client's fail-secure
sandboxing check. The protocol exposes filesystem writes without protocol-level
restrictions on target paths, so an agent may modify
security-sensitive locations outside the intended project scope.
Because permission is \textbf{may}-level rather than mandatory,
this check can be skipped entirely; combined with content-integrity
gaps, untrusted repository content can steer the agent toward persistent
security-critical changes. More broadly, security architecture does
not imply assurance. ANP includes the strongest built-in
mechanisms in our study---DID-based identity, end-to-end
encryption, and token lifecycle support---yet still fails on
consent, audit, fail-secure replay prevention, and credential
revocation. 

\textbf{Representative counterexamples.}\label{sec:eval:cex}
We highlight three representative counterexamples. Each illustrates a
different binding between a failed invariant and protocol evidence:
recommendation gap, framework-derived design gap, and standards
non-conformance.
(1)~\textbf{MCP credential/registry integrity}
(\textbf{spec-recommended}). TLC finds a trace where a session is
initialized and closed, yet the credential is never revoked. No
adversary action is required: session termination does not terminate
the authority that the session created. The spec does not specify
revocation and only recommends token lifetime controls, so this is a
recommendation/lifecycle gap rather than standards non-conformance.
(2)~\textbf{A2A delegation monotonicity}
(\textbf{framework hardening}). TLC finds a trace where one agent
delegates a capability, after which an adversarial step re-delegates
it beyond the receiving agent's original grant. The spec does not
define delegation-scope constraints, so transitive re-delegation can
amplify authority across hops. This is a framework-derived design gap, not a
violation of an explicit A2A requirement.
(3)~\textbf{MCP content integrity} (\textbf{spec-mandated}). TLC
finds a trace where a compromised tool response is incorporated into
the system's control state---a protocol-level prompt-injection
failure. The MCP spec explicitly states that servers \textbf{must}
sanitize tool outputs, but the reference SDK performs zero
sanitization, making this a direct standards non-conformance.

\subsection{RQ2: Implementation Reproduction}\label{sec:eval:impl}

Phase~2 uses 80~implementation tests: 18~MCP SDK tests, 13~A2A
tests, 16~ACP-Client tests, 8~ANP tests, 9~ACP-Cap tests,
9~composition tests, and 7~MCP reference-server proof-of-concept
tests. Table~\ref{tab:impl-results} reports the outcomes. We
separate \emph{behavior claims} from \emph{absence claims}:
behavioral tests show that an executable SDK path reaches a
violating state, while source/type-level checks support claims that
no callable hook, endpoint, field, or type exists to enforce the
property. This distinction is why MCP has the strongest evidence
(SDK, composition, and reference-server tests), A2A has behavioral
SDK evidence but no server proof-of-concept target, and composition
pairs without production bridges remain model-only.

\begin{table}[t]
\centering
\caption{Implementation-level results summary
Coverage: B~=~includes behavioral evidence, S~=~source/type-only
absence or structural evidence.}
\label{tab:impl-results}

\footnotesize
\setlength{\tabcolsep}{2.5pt}
\renewcommand{\arraystretch}{1.08}

\begin{tabularx}{\columnwidth}{@{}
  >{\raggedright\arraybackslash}p{0.23\columnwidth}
  >{\centering\arraybackslash}p{0.08\columnwidth}
  >{\centering\arraybackslash}p{0.07\columnwidth}
  >{\raggedright\arraybackslash}X
@{}}
\toprule
\textbf{Target} & \textbf{Tests} & \textbf{Cov.} & \textbf{Outcome summary} \\
\midrule
MCP        & 18 & B & capability, content, consent, audit, credential = BOTH; fail-secure = SPEC \\
A2A        & 13 & B & delegation, consent, audit, credential = BOTH \\
ACP-Client & 16 & B & capability, content, consent, audit, fail-secure, credential = BOTH \\
ANP        &  8 & B & identity, format = pass; consent, audit = BOTH; fail-secure, credential = partial$^\ddagger$ \\
ACP-Cap    &  9 & B & format, session = pass; consent, audit, credential = BOTH; fail-secure = partial$^\ddagger$ \\

\midrule
\multicolumn{4}{@{}l}{\emph{Composition harness}} \\
MCP+MCP chained       & 4 & B & 4 composition findings \\
MCP+ACP-Cl composed   & 4 & S & 4 composition findings \\
MCP+ACP-Cl behavioral & 1 & B & injection confirmed \\

\midrule
\multicolumn{4}{@{}l}{\emph{MCP reference-server proof-of-concept tests$^\ast$}} \\
fetch  & 2 & B & 5/5 web payloads \\
sqlite & 2 & B & 3/3 stored payloads \\
git    & 3 & B & 3/3 vectors \\

\midrule
\textbf{Total} & \textbf{80} & & \\
\bottomrule
\end{tabularx}

\smallskip
\raggedright\scriptsize
$^\ast$Proof-of-concept counts are distinct injection patterns that passed
through the server unsanitized.
$^\ddagger$``partial'' = mechanism exists in the SDK but its
enforcement is incomplete.
\end{table}

\textbf{Key results.}
Nine of the ten MCP$\leftrightarrow$A2A behavioral findings reproduce at both the
specification and implementation levels; the exception is MCP's
fail-secure/session check, where the HTTP transport layer compensates
for the session-layer gap. MCP's content-integrity check is a direct
non-conformance: the specification says
servers \textbf{must} sanitize tool outputs, but five injection
payloads pass through the SDK tool-call path unmodified. By
contrast, A2A's delegation, consent, audit, and credential-lifecycle
checks are framework hardening gaps because the specification does
not require delegation scope, consent, audit, or credential
revocation. The ANP fail-secure/credential checks and
ACP-Cap fail-secure checks only provide partial mechanisms.

\subsection{RQ3: Composition Failures}\label{sec:eval:composition}

We now evaluate the eight composed models constructed via the
composition methodology of \S\ref{sec:framework:composition}.
Across these models, \sys{} instantiates 43~security-oriented
composition obligations: 31~boundary-risk obligations and
12~bridge-contract preservation obligations. Thirty admit
counterexamples (Table~\ref{tab:cs-results}). We also check
40~bounded model-validation invariants; all pass and are excluded
from this denominator because they validate the encoding rather than
security obligations. Thus, 30/43 is the counterexample rate for
layer-derived cross-boundary security obligations, not for all TLC
checks.

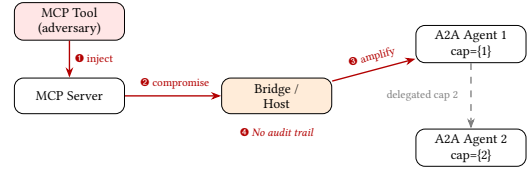
\begin{figure}[t]
\centering
\resizebox{0.8\columnwidth}{!}{%
\begin{tikzpicture}[
  box/.style={
    rectangle, draw, rounded corners,
    minimum width=1.7cm,
    minimum height=0.55cm,
    inner sep=2pt,
    align=center,
    font=\scriptsize
  },
  attack/.style={-{Stealth[length=1.6mm]}, semithick, red!70!black},
  normal/.style={-{Stealth[length=1.6mm]}, semithick, dashed, gray},
  label/.style={font=\tiny},
]

\node[box, fill=red!10] (tool) at (0, 0) {MCP Tool\\(adversary)};
\node[box] (mcp) at (0, -1.15) {MCP Server};

\node[box, fill=orange!15] (bridge) at (3.2, -1.15) {Bridge /\\Host};

\node[box] (ag1) at (6.2, -0.35) {A2A Agent 1\\cap=\{1\}};
\node[box] (ag2) at (6.2, -1.95) {A2A Agent 2\\cap=\{2\}};

\draw[attack] (tool) -- (mcp);
\node[label, text=red!70!black] at (0.38, -0.58) {\ding{202} inject};

\draw[attack] (mcp) -- (bridge)
  node[pos=0.52, above, font=\tiny, text=red!70!black] {\ding{203} compromise};

\draw[attack] (bridge) -- (ag1)
  node[pos=0.52, above, sloped, font=\tiny, text=red!70!black] {\ding{204} amplify};

\draw[normal] (ag1) -- (ag2);
\node[font=\tiny, text=gray] at (5.45, -1.15) {delegated cap 2};

\node[font=\tiny\itshape, text=red!70!black, below=0.10cm of bridge]
  {\ding{205} No audit trail};

\end{tikzpicture}%
}
\caption{Cross-protocol attack chain: MCP prompt injection propagates through a shared bridge to amplify A2A delegation.}
\label{fig:composition-attack}
\vspace{-0.5em}
\end{figure}

\textbf{Representative case study: MCP$\leftrightarrow$A2A.}
Figure~\ref{fig:composition-attack} shows a common composed
deployment: MCP tool output enters an A2A conductor that coordinates
downstream agents. The composed model keeps the MCP session and
tool-output state, the A2A capability and delegation state, and the
bridge action that moves content between them. TLC finds a trace
where unsanitized MCP output taints the bridge; the bridge then
causes A2A authority to exceed the recipient's original grant, and
the escalation has no complete audit trail because the bridge sits
outside both local audit scopes. This failure is not visible in
either protocol alone: MCP exposes a content-integrity gap and A2A
exposes a delegation-scope gap, but the authority expansion appears
only at the boundary.

\textbf{Results across protocol pairs.}
We extend beyond the MCP$\leftrightarrow$A2A case study to the broader space of
realistic protocol pairings. Table~\ref{tab:cs-results}
summarizes the results: 30~counterexamples across
43~security-oriented composition obligations in 8~composed models.

\begin{table}[t]
\centering
\caption{Security-oriented composition obligations
across 8~composed models. CEX = counterexamples.}
\label{tab:cs-results}
\small
\setlength{\tabcolsep}{3pt}
\resizebox{\columnwidth}{!}{%
\begin{tabular}{@{}lcccl@{}}
\toprule
\textbf{Model} & \textbf{Sec. oblig.} & \textbf{CEX} &
\textbf{Layers} & \textbf{Representative finding} \\
\midrule
MCP$\leftrightarrow$A2A             & 9 & 7 & L2--L5 & bridge-driven authority escalation \\
MCP$\leftrightarrow$MCP chained     & 7 & 5 & L3--L5 & injection chain / transitive trust \\
MCP$\leftrightarrow$ACP-Cap         & 4 & 3 & L3--L5 & consent bypass \\
A2A$\leftrightarrow$ACP-Cap         & 5 & 3 & L2--L3 & authority-model mismatch \\
A2A$\leftrightarrow$A2A federated   & 4 & 2 & L3 & federation amplification \\
MCP$\leftrightarrow$ACP-Client      & 6 & 4 & L3--L5 & injection-to-action flow \\
ANP$\leftrightarrow$MCP             & 4 & 3 & L3--L5 & identity $\neq$ content safety \\
ANP$\leftrightarrow$A2A             & 4 & 3 & L3/L5 & identity $\neq$ delegation control \\
\midrule
\textbf{Total}      & \textbf{43} & \textbf{30} & & \\
\bottomrule
\end{tabular}
}
\end{table}

\textbf{Cross-pair synthesis.}
The seven additional protocol pairings cluster into three failure
modes that recur regardless of which two protocols we compose.
(1) \textbf{Hidden intermediaries:} chained MCP relays, A2A federation,
and ANP-discovered MCP tool use all produce failed checks that
neither endpoint's local checks can observe (e.g., transitive
credential cascade, federated delegation amplification, encrypted
injection). (2) \textbf{Authority/consent semantic mismatches:}
MCP$\leftrightarrow$ACP-Cap and A2A$\leftrightarrow$ACP-Cap fail because consent and delegation
models differ across the two protocols and the bridge resolves
the difference silently. (3) \textbf{Identity is orthogonal to content
and scope:} the two ANP compositions confirm that even DID-based
identity verification provides no mechanism to bound content
integrity or delegation scope.

\textbf{Implications.}
The composition results support four conclusions.
(1)~\textbf{Individual protocol analysis is necessary but
insufficient.} A protocol can satisfy its own local checks and
still fail once its outputs become another protocol's trusted
inputs.
(2)~\textbf{Composition failures are heterogeneous but systematic.}
Some are authority-amplification failures, some are
consent-routing failures, some are provenance and audit failures,
and some are direct content-to-action flows. What unifies them is
that they emerge only at the boundary between protocols.
(3)~\textbf{Two structural patterns recur: hiding intermediaries
and exposing local actions.} Chained MCP collapses provenance,
trust, and credential boundaries in a common deployment pattern.
MCP$\leftrightarrow$ACP-Client turns untrusted semantic content into local side
effects.
(4)~\textbf{Bridge components are unowned responsibility points.} The most
security-critical logic often sits in the bridge, conductor,
relay, or agent runtime that connects two protocols, yet these
components are typically outside the normative security scope of
either protocol. Composition safety is therefore not only a
verification problem, but also a responsibility-allocation problem.

\subsection{RQ4: Responsibility Allocation}\label{sec:eval:responsibility}

We apply the Responsibility IR methodology of
\S\ref{sec:framework:responsibility-ir} to obtain 35 responsibility
records across four protocols (MCP~14, A2A~9, ANP~5,
ACP-Client~7; ACP-Cap is archived). The findings expose who owns
each security control and whether that ownership translates into
enforcement.

\begin{figure}[t]
\centering
\scriptsize
\setlength{\tabcolsep}{2.2pt}
\renewcommand{\arraystretch}{1.1}
\resizebox{0.9\columnwidth}{!}{%
\begin{tabular}{l|ccccccccccc}
\toprule
& \rotatebox{70}{Id.}
& \rotatebox{70}{Cap.}
& \rotatebox{70}{Deleg.}
& \rotatebox{70}{Content}
& \rotatebox{70}{Consent}
& \rotatebox{70}{Audit}
& \rotatebox{70}{Fail-sec.}
& \rotatebox{70}{Cred.}
& \rotatebox{70}{Format}
& \rotatebox{70}{Session}
& \rotatebox{70}{Comp.} \\
\midrule
MCP
& \cellcolor{red!30}A
& \cellcolor{orange!30}C
& ---
& \cellcolor{purple!30}D
& \cellcolor{orange!30}C
& \cellcolor{orange!30}C
& \cellcolor{yellow!30}B
& \cellcolor{orange!30}C
& \cellcolor{green!20}\ding{51}
& \cellcolor{green!20}\ding{51}
& \cellcolor{blue!20}E \\
A2A
& \cellcolor{green!20}\ding{51}
& \cellcolor{orange!30}C
& \cellcolor{red!30}A
& \cellcolor{red!30}A
& \cellcolor{orange!30}C
& \cellcolor{red!30}A
& \cellcolor{orange!30}C
& \cellcolor{orange!30}C
& \cellcolor{green!20}\ding{51}
& \cellcolor{green!20}\ding{51}
& \cellcolor{blue!20}E \\
ANP
& \cellcolor{green!20}\ding{51}
& ---
& ---
& ---
& \cellcolor{orange!30}C
& \cellcolor{red!30}A
& \cellcolor{green!20}\ding{51}
& \cellcolor{orange!30}C
& \cellcolor{green!20}\ding{51}
& \cellcolor{green!20}\ding{51}
& --- \\
ACP-Client
& \cellcolor{red!30}A
& \cellcolor{red!30}A
& ---
& \cellcolor{red!30}A
& \cellcolor{orange!30}C
& \cellcolor{red!30}A
& \cellcolor{orange!30}C
& \cellcolor{red!30}A
& \cellcolor{green!20}\ding{51}
& \cellcolor{green!20}\ding{51}
& \cellcolor{blue!20}E \\
\bottomrule
\end{tabular}
}

\vspace{4pt}
{\scriptsize
\colorbox{red!30}{A}~Ownership gap \quad
\colorbox{yellow!30}{B}~Split-duty \quad
\colorbox{orange!30}{C}~Enforcement gap \quad
\colorbox{purple!30}{D}~Scope gap \quad
\colorbox{blue!20}{E}~Composition control \quad
\colorbox{green!20}{\ding{51}}~Enforced \quad
---~N/A
}
\caption{Responsibility gap heatmap across layer-derived checks.
Agent-layer and composition controls are
dominated by ownership, enforcement, scope, and orphan gaps.}
\label{fig:responsibility-heatmap}
\end{figure}

\textbf{Key results.}
Figure~\ref{fig:responsibility-heatmap} shows three patterns. First,
only one security-relevant control is both owned and enforced: ANP's
DID-based identity verification check; the other enforced controls
are structural wire-format or session-lifecycle checks. Second,
enforcement gaps dominate: 18 of 32 total gaps assign responsibility
in the specification but provide no SDK mechanism. Third,
composition is universally orphaned: no protocol owns bridge
security for chained MCP servers, A2A conductors, or MCP tools
inside ACP-Client agents. These results explain why local protocol
fixes are insufficient: the most security-critical obligations often
belong to the runtime bridge or deployment policy that connects two
otherwise independent protocols.

\section{Discussion}\label{sec:discussion}

\textbf{Why composition failures are hard to fix}.
Single-protocol findings usually have a clear disclosure target:
the specification author, the SDK maintainer, or the application
developer. Cross-protocol composition findings do not. In cases
such as MCP$\leftrightarrow$ACP-Client, sanitization, local-action control, and
cross-protocol behavior are allocated across different parties, but
no single party controls the full composed path. This is why
composition failures often survive the normal disclosure pipeline:
the responsibility chain has no clear endpoint. Our results suggest
that composition safety should be treated as a first-class protocol
property, not as an application-level afterthought.

\textbf{Implications for protocol deployment}.
The broader lesson is that agent protocol security cannot be
improved solely through better individual specifications or SDK
patches. It also requires coordination mechanisms for security
requirements that span protocol boundaries. Mature communication
stacks have such structures; agent protocols largely do not.
Our contribution is therefore not only to identify concrete gaps,
but to provide protocol-derived checks and a source-linked
analysis workflow that makes those cross-protocol requirements
explicit early in the design process.

\section{Related Work}\label{sec:related}

\textbf{Agentic software systems and protocol infrastructure.}
Recent work on LLM and agent security has studied prompt injection,
tool abuse, agent benchmarks, and deployment-level threat
models~\cite{greshake2023youve,zhan2024injecagent,debenedetti2024agentdojo,299563,liu2025wainjectbench0,anbiaee2026security,kong2025survey,DBLP:journals/csur/HeZYLZY26}.
Developer-facing discussions increasingly treat MCP, A2A, ANP, ACP,
and ACP-Client as infrastructure for tool use, coding agents, and
multi-agent workflows~\cite{mcp2024,a2a2024,anp2025,acp2025,acpclient2025,saboo2026devguide}.
Qualitative work has begun to analyze A2A--MCP integration as a
glue-code-to-protocol problem~\cite{li2025glue}, and security labs
have reported concrete failures in MCP server deployments
~\cite{snyk2025mcp}. \sys{} complements this work by treating the
protocol-mediated deployment stack---specification, SDK, runtime
bridge, and responsibility allocation---as the unit of analysis.

\textbf{Formal verification and conformance testing.}
Security has long been integral to protocol design~\cite{rfc3552},
with formal analysis applied to classical and deployed protocols
via CSP~\cite{lowe1996breaking}, ProVerif and
Tamarin~\cite{blanchet2001proverif,meier2013tamarin}, and direct
reasoning against modern standards
~\cite{basin2018formal,bhargavan2002formal,10.1145/3522582,287366,287101,299715,255320,285435,10.5555/3766078.3766441,10.5555/3766078.3766336}.
Agentic formal-methods systems address adjacent objects: VeriPlan
verifies end-user plans~\cite{lee2025veriplan}, and TraceFix repairs
coordination logic from \tlaplus{} counterexamples
~\cite{xia2026tracefix}. \sys{} instead checks public communication
protocols and replays counterexamples against SDKs, following the
tradition of protocol conformance testing
~\cite{bishop2005rigorous,10.1109/49.44561,10.1145/186258.187153,10.1145/18172.18199}
and verification-linked implementation assurance
~\cite{10.1145/3576915.3623105}.

\textbf{Specification extraction and responsibility evidence.}
Our extraction front end builds on work that recovers behavioral
artifacts from protocol prose, including state machines, parseable
structure, ambiguity, and parser constraints
~\cite{pacheco2022,hermes2024,prosper2023,li2025extracting,zheng2025parcleanse,sage2021,basin2025bridging}.
The closest control-evidence analogue is OSCAL~\cite{oscal}, which
structures security controls and evidence. Our Responsibility IR
adapts that idea to protocol-derived controls and connects it to
system-security and requirements-analysis concerns
~\cite{279972,10.1145/3172871.3172879}: when agent protocols are
composed, the missing requirement is often not a local check but an
unowned bridge responsibility.

\section{Threats to Validity}\label{sec:threats}

\textbf{Construct validity.}
Our layer-derived checks may not always correspond to explicit
design goals of every protocol. We therefore classify failures by evidence: only MUST-backed failures
are standards nonconformance; SHOULD-backed failures are
recommendation gaps; and framework or layer-completeness failures
are design or responsibility gaps.

\textbf{Modeling and extraction validity.}
Protocol documents can be ambiguous, incomplete, or inconsistent
with their SDKs. Although clause extraction and typed-IR preparation
are validation-gated, the resulting IR is still an abstraction of
the analyzed source snapshots. To reduce this threat, records retain
source references and modality tags, ambiguous clauses are modeled
as nondeterministic alternatives, generated \tlaplus{} actions are
type-checked, and executable counterexamples are replayed when an
implementation surface exists. We model protocol effects of semantic
content and delegated authority, not full LLM internals.

\textbf{External validity.}
Our empirical results cover five protocols, selected SDKs and
reference servers, bounded TLC configurations, and the protocol
versions available during the study. The results should not be read
as prevalence estimates or proofs for unbounded systems. We mitigate
this by choosing bounds that instantiate observed attack schemas,
separating behavioral from source/type evidence, and versioning
source snapshots and generated artifacts.

\section{Conclusion}\label{sec:conclusion}

Agent protocols are becoming the connective tissue of agentic
software systems, but their security properties are split across
specifications, SDKs, runtime bridges, and application policy.
\sys{} makes these split obligations explicit by turning
protocol sources into formal checks, counterexample traces,
implementation tests, and responsibility records. Across five
protocols and eight composed models, we find that many failures
appear only when semantic content, delegated authority, and tool
access cross protocol boundaries. Future protocols therefore need
not only stronger local requirements, but also explicit composition
contracts and enforcement responsibility for the runtimes that join
them.

\bibliographystyle{ACM-Reference-Format}
\bibliography{references}

@misc{mcp2024,
  title = {Model Context Protocol Specification},
  author = {Anthropic},
  year = {2024},
  howpublished = {\url{https://modelcontextprotocol.io/specification/2025-11-25}},
  note = {retrieved March 2026}
}

@misc{a2a2024,
  title = {Agent-to-Agent Protocol ({A2A})},
  author = {{A2A Project}},
  year = {2025},
  howpublished = {\url{https://github.com/a2aproject/A2A}},
  note = {Announced April 2025; now under Linux Foundation. Retrieved March 2026}
}

@misc{anp2025,
  title = {{AgentNetworkProtocol} ({ANP}): An Open Protocol for Agent Communication},
  author = {{Agent Network Protocol Team}},
  year = {2025},
  howpublished = {\url{https://github.com/agent-network-protocol/AgentNetworkProtocol}},
  note = {Retrieved March 2026}
}

@misc{acp2025,
  title = {Agent Communication Protocol},
  author = {{ACP Contributors}},
  year = {2025},
  howpublished = {\url{https://agentcommunicationprotocol.dev/introduction/welcome}},
  note = {Under active development via RFDs. Retrieved March 2026}
}

@book{lamport2002specifying,
  title={Specifying systems},
  author={Lamport, Leslie},
  volume={388},
  year={2002},
  publisher={Addison-Wesley Boston}
}

@inproceedings{yu1999model,
  title={Model checking TLA+ specifications},
  author={Yu, Yuan and Manolios, Panagiotis and Lamport, Leslie},
  booktitle={Advanced research working conference on correct hardware design and verification methods},
  pages={54--66},
  year={1999},
  organization={Springer}
}

@article{dolev1983security,
  title={On the security of public key protocols},
  author={Dolev, Danny and Yao, Andrew},
  journal={IEEE Transactions on information theory},
  volume={29},
  number={2},
  pages={198--208},
  year={2003},
  publisher={IEEE}
}

@inproceedings{lowe1996breaking,
  title={Breaking and fixing the Needham-Schroeder public-key protocol using FDR},
  author={Lowe, Gavin},
  booktitle={International Workshop on Tools and Algorithms for the Construction and Analysis of Systems},
  pages={147--166},
  year={1996},
  organization={Springer}
}

@inproceedings{blanchet2001proverif,
  title={An efficient cryptographic protocol verifier based on prolog rules. 2014},
  author={Blanchet, B and others},
  journal={doi},
  volume={10},
  pages={82--96}
}

@inproceedings{meier2013tamarin,
  title={The TAMARIN prover for the symbolic analysis of security protocols},
  author={Meier, Simon and Schmidt, Benedikt and Cremers, Cas and Basin, David},
  booktitle={International conference on computer aided verification},
  pages={696--701},
  year={2013},
  organization={Springer}
}

@inproceedings{basin2018formal,
    author = {Basin, David and Dreier, Jannik and Hirschi, Lucca and Radomirovic, Sa\v{s}a and Sasse, Ralf and Stettler, Vincent},
    title = {A Formal Analysis of 5G Authentication},
    year = {2018},
    isbn = {9781450356930},
    publisher = {Association for Computing Machinery},
    address = {New York, NY, USA},
    url = {https://doi.org/10.1145/3243734.3243846},
    doi = {10.1145/3243734.3243846},
    booktitle = {Proceedings of the 2018 ACM SIGSAC Conference on Computer and Communications Security},
    pages = {1383–1396},
    numpages = {14},
    location = {Toronto, Canada},
    series = {CCS '18}
}

@inproceedings{greshake2023youve,
  title={Not what you've signed up for: Compromising real-world llm-integrated applications with indirect prompt injection},
  author={Greshake, Kai and Abdelnabi, Sahar and Mishra, Shailesh and Endres, Christoph and Holz, Thorsten and Fritz, Mario},
  booktitle={Proceedings of the 16th ACM workshop on artificial intelligence and security},
  pages={79--90},
  year={2023}
}

@inproceedings{zhan2024injecagent,
  title={Injecagent: Benchmarking indirect prompt injections in tool-integrated large language model agents},
  author={Zhan, Qiusi and Liang, Zhixiang and Ying, Zifan and Kang, Daniel},
  booktitle={Findings of the Association for Computational Linguistics: ACL 2024},
  pages={10471--10506},
  year={2024}
}

@inproceedings{debenedetti2024agentdojo,
  title={Agentdojo: A dynamic environment to evaluate prompt injection attacks and defenses for llm agents},
  author={Debenedetti, Edoardo and Zhang, Jie and Balunovic, Mislav and Beurer-Kellner, Luca and Fischer, Marc and Tram{\`e}r, Florian},
  journal={Advances in Neural Information Processing Systems},
  volume={37},
  pages={82895--82920},
  year={2024}
}

@inproceedings{pacheco2022,
  title={Automated attack synthesis by extracting finite state machines from protocol specification documents},
  author={Pacheco, Maria Leonor and von Hippel, Max and Weintraub, Ben and Goldwasser, Dan and Nita-Rotaru, Cristina},
  booktitle={2022 IEEE Symposium on Security and Privacy (SP)},
  pages={51--68},
  year={2022},
  organization={IEEE}
}

@inproceedings{hermes2024,
  title={Hermes: Unlocking security analysis of cellular network protocols by synthesizing finite state machines from natural language specifications},
  author={Al Ishtiaq, Abdullah and Das, Sarkar Snigdha Sarathi and Rashid, Syed Md Mukit and Ranjbar, Ali and Tu, Kai and Wu, Tianwei and Song, Zhezheng and Wang, Weixuan and Akon, Mujtahid and Zhang, Rui and others},
  booktitle={33rd USENIX Security Symposium (USENIX Security 24)},
  pages={4445--4462},
  year={2024}
}

@inproceedings{prosper2023,
  title={Prosper: Extracting protocol specifications using large language models},
  author={Sharma, Prakhar and Yegneswaran, Vinod},
  booktitle={Proceedings of the 22nd ACM workshop on hot topics in networks},
  pages={41--47},
  year={2023}
}

@inproceedings{bishop2005rigorous,
    author = {Bishop, Steve and Fairbairn, Matthew and Norrish, Michael and Sewell, Peter and Smith, Michael and Wansbrough, Keith},
    title = {Rigorous specification and conformance testing techniques for network protocols, as applied to TCP, UDP, and sockets},
    year = {2005},
    issue_date = {October 2005},
    publisher = {Association for Computing Machinery},
    address = {New York, NY, USA},
    volume = {35},
    number = {4},
    issn = {0146-4833},
    url = {https://doi.org/10.1145/1090191.1080123},
    doi = {10.1145/1090191.1080123},
    month = aug,
    pages = {265–276},
    numpages = {12}
}

@article{bhargavan2002formal,
    author = {Bhargavan, Karthikeyan and Obradovic, Davor and Gunter, Carl A.},
    title = {Formal verification of standards for distance vector routing protocols},
    year = {2002},
    issue_date = {July 2002},
    publisher = {Association for Computing Machinery},
    address = {New York, NY, USA},
    volume = {49},
    number = {4},
    issn = {0004-5411},
    url = {https://doi.org/10.1145/581771.581775},
    doi = {10.1145/581771.581775},
    journal = {J. ACM},
    month = jul,
    pages = {538–576},
    numpages = {39}
}

@inproceedings{li2025extracting,
  title={Extracting Formal Specifications From Documents Using LLMS for Test Automation},
  author={Li, Hui and Dong, Zhen and Wang, Siao and Zhang, Hui and Shen, Liwei and Peng, Xin and She, Dongdong},
  booktitle={2025 IEEE/ACM 33rd International Conference on Program Comprehension (ICPC)},
  pages={1--12},
  year={2025},
  organization={IEEE Computer Society}
}

@inproceedings{zheng2025parcleanse,
    author = {Zheng, Mingwei and Xie, Danning and Shi, Qingkai and Wang, Chengpeng and Zhang, Xiangyu},
    title = {Validating Network Protocol Parsers with Traceable RFC Document Interpretation},
    year = {2025},
    issue_date = {July 2025},
    publisher = {Association for Computing Machinery},
    address = {New York, NY, USA},
    volume = {2},
    number = {ISSTA},
    url = {https://doi.org/10.1145/3728955},
    doi = {10.1145/3728955},
    journal = {Proc. ACM Softw. Eng.},
    month = jun,
    articleno = {ISSTA078},
    numpages = {23}
}

@inproceedings{sage2021,
    author = {Yen, Jane and L\'{e}vai, Tam\'{a}s and Ye, Qinyuan and Ren, Xiang and Govindan, Ramesh and Raghavan, Barath},
    title = {Semi-automated protocol disambiguation and code generation},
    year = {2021},
    isbn = {9781450383837},
    publisher = {Association for Computing Machinery},
    address = {New York, NY, USA},
    url = {https://doi.org/10.1145/3452296.3472910},
    doi = {10.1145/3452296.3472910},
    pages = {272–286},
    numpages = {15},
    location = {Virtual Event, USA},
    series = {SIGCOMM '21}
}

@article{basin2025bridging,
    author = {Basin, David and Foster, Nate and McMillan, Kenneth L. and Namjoshi, Kedar S. and Nita-Rotaru, Cristina and Smith, Jonathan M. and Zave, Pamela and Zuck, Lenore D.},
    title = {It Takes a Village: Bridging the Gaps between Current and Formal Specifications for Protocols},
    year = {2025},
    issue_date = {August 2025},
    publisher = {Association for Computing Machinery},
    address = {New York, NY, USA},
    volume = {68},
    number = {8},
    issn = {0001-0782},
    url = {https://doi.org/10.1145/3706572},
    doi = {10.1145/3706572},
    journal = {Commun. ACM},
    month = jul,
    pages = {50–61},
    numpages = {12}
}

@misc{rfc2119,
  title = {Key words for use in {RFC}s to Indicate Requirement Levels},
  author={Bradner, Scott},
  howpublished = {RFC 2119},
  year = {1997}
}

@misc{rfc3552,
  title = {Guidelines for Writing {RFC} Text on Security Considerations},
  author={Rescorla, Eric and Korver, Brian},
  howpublished = {RFC 3552},
  year = {2003}
}

@misc{rfc5531,
  title = {{RPC}: Remote Procedure Call Protocol Specification Version 2},
  author = {Thurlow, Robert},
  howpublished = {RFC 5531},
  year = {2009}
}

@misc{rfc7861,
  title = {Remote Procedure Call ({RPC}) Security Version 3},
  author = {Adamson, William and Williams, Nicolas},
  howpublished = {RFC 7861},
  year = {2016}
}

@misc{cve202549596,
  title = {{CVE}-2025-49596: {MCP Inspector} remote code execution vulnerability},
  author = {{National Vulnerability Database}},
  year = {2025},
  howpublished = {\url{https://nvd.nist.gov/vuln/detail/CVE-2025-49596}}
}

@misc{cve202568143,
  title = {{CVE}-2025-68143: {Anthropic} {MCP} Git server path handling vulnerability},
  author = {{National Vulnerability Database}},
  year = {2025},
  howpublished = {\url{https://nvd.nist.gov/vuln/detail/CVE-2025-68143}}
}

@misc{cve202568144,
  title = {{CVE}-2025-68144: {Anthropic} {MCP} Git server argument injection vulnerability},
  author = {{National Vulnerability Database}},
  year = {2025},
  howpublished = {\url{https://nvd.nist.gov/vuln/detail/CVE-2025-68144}}
}

@misc{cve202568145,
  title = {{CVE}-2025-68145: {Anthropic} {MCP} Git server repository boundary vulnerability},
  author = {{National Vulnerability Database}},
  year = {2025},
  howpublished = {\url{https://nvd.nist.gov/vuln/detail/CVE-2025-68145}}
}

@misc{acpclient2025,
  title = {Agent Client Protocol ({ACP})},
  author = {{Agent Client Protocol Contributors}},
  year = {2025},
  howpublished = {\url{https://github.com/agentclientprotocol/agent-client-protocol}},
  note = {retrieved March 2026}
}

@article{anbiaee2026security,
  title   = {Security Threat Modeling for Emerging AI-Agent Protocols: A Comparative Analysis of MCP, A2A, Agora, and ANP},
  author  = {Zeynab Anbiaee and Mahdi Rabbani and Mansur Mirani and Gunjan Piya and Igor Opushnyev and Ali Ghorbani and Sajjad Dadkhah},
  year    = {2026},
  journal = {arXiv preprint arXiv: 2602.11327}
}

@article{kong2025survey,
  title   = {A Survey of LLM-Driven AI Agent Communication: Protocols, Security Risks, and Defense Countermeasures},
  author  = {Dezhang Kong and Shi Lin and Zhenhua Xu and Zhebo Wang and Minghao Li and Yufeng Li and Yilun Zhang and Hujin Peng and Xiang Chen and Zeyang Sha and Yuyuan Li and Changting Lin and Xun Wang and Xuan Liu and Ningyu Zhang and Chaochao Chen and Chunming Wu and Muhammad Khurram Khan and Meng Han},
  year    = {2025},
  journal = {arXiv preprint arXiv: 2506.19676}
}

@article{DBLP:journals/csur/HeZYLZY26,
  author    = {Feng He and Tianqing Zhu and Dayong Ye and Bo Liu and Wanlei Zhou and Philip S. Yu},
  title     = {The Emerged Security and Privacy of {LLM} Agent: {A} Survey with Case Studies},
  journal   = {{ACM} Comput. Surv.},
  volume    = {58},
  number    = {6},
  pages     = {162:1-162:36},
  year      = {2026},
  url       = {https://doi.org/10.1145/3773080},
  doi       = {10.1145/3773080},
  bibsource = {dblp computer science bibliography, https://dblp.org}
}

@inproceedings {299563,
author = {Yupei Liu and Yuqi Jia and Runpeng Geng and Jinyuan Jia and Neil Zhenqiang Gong},
title = {Formalizing and Benchmarking Prompt Injection Attacks and Defenses},
booktitle = {33rd USENIX Security Symposium (USENIX Security 24)},
year = {2024},
isbn = {978-1-939133-44-1},
address = {Philadelphia, PA},
pages = {1831--1847},
url = {https://www.usenix.org/conference/usenixsecurity24/presentation/liu-yupei},
publisher = {USENIX Association},
month = aug
}

@article{liu2025wainjectbench0,
  title   = {WAInjectBench: Benchmarking Prompt Injection Detections for Web Agents},
  author  = {Yinuo Liu and Ruohan Xu and Xilong Wang and Yuqi Jia and Neil Zhenqiang Gong},
  year    = {2025},
  journal = {arXiv preprint arXiv: 2510.01354}
}

@article{10.1145/3522582,
author = {Kulik, Tomas and Dongol, Brijesh and Larsen, Peter Gorm and Macedo, Hugo Daniel and Schneider, Steve and Tran-J\o{}rgensen, Peter W. V. and Woodcock, James},
title = {A Survey of Practical Formal Methods for Security},
year = {2022},
issue_date = {March 2022},
publisher = {Association for Computing Machinery},
address = {New York, NY, USA},
volume = {34},
number = {1},
issn = {0934-5043},
url = {https://doi.org/10.1145/3522582},
doi = {10.1145/3522582},
journal = {Form. Asp. Comput.},
month = jul,
articleno = {5},
numpages = {39}
}

@inproceedings {299715,
author = {Jacob Ginesin and Max von Hippel and Evan Defloor and Cristina Nita-Rotaru and Michael T{\"u}xen},
title = {A Formal Analysis of {SCTP}: Attack Synthesis and Patch Verification},
booktitle = {33rd USENIX Security Symposium (USENIX Security 24)},
year = {2024},
isbn = {978-1-939133-44-1},
address = {Philadelphia, PA},
pages = {3099--3116},
url = {https://www.usenix.org/conference/usenixsecurity24/presentation/ginesin},
publisher = {USENIX Association},
month = aug
}

@inproceedings{10.1145/3576915.3623105,
author = {Arquint, Linard and Schwerhoff, Malte and Mehta, Vaibhav and M\"{u}ller, Peter},
title = {A Generic Methodology for the Modular Verification of Security Protocol Implementations},
year = {2023},
isbn = {9798400700507},
publisher = {Association for Computing Machinery},
address = {New York, NY, USA},
url = {https://doi.org/10.1145/3576915.3623105},
doi = {10.1145/3576915.3623105},
booktitle = {Proceedings of the 2023 ACM SIGSAC Conference on Computer and Communications Security},
pages = {1377–1391},
numpages = {15},
location = {Copenhagen, Denmark},
series = {CCS '23}
}

@article{10.1109/49.44561,
author = {Linn, R. J.},
title = {Conformance evaluation methodology and protocol testing},
year = {2006},
issue_date = {September 2006},
publisher = {IEEE Press},
volume = {7},
number = {7},
issn = {0733-8716},
url = {https://doi.org/10.1109/49.44561},
doi = {10.1109/49.44561},
journal = {IEEE J.Sel. A. Commun.},
month = sep,
pages = {1143–1158},
numpages = {16}
}

@inproceedings{10.1145/186258.187153,
author = {Bochmann, Gregor V. and Petrenko, Alexandre},
title = {Protocol testing: review of methods and relevance for software testing},
year = {1994},
isbn = {0897916832},
publisher = {Association for Computing Machinery},
address = {New York, NY, USA},
url = {https://doi.org/10.1145/186258.187153},
doi = {10.1145/186258.187153},
booktitle = {Proceedings of the 1994 ACM SIGSOFT International Symposium on Software Testing and Analysis},
pages = {109–124},
numpages = {16},
location = {Seattle, Washington, USA},
series = {ISSTA '94}
}

@inproceedings{10.1145/18172.18199,
author = {Sarikaya, B},
title = {Formal specification-based conformance testing},
year = {1986},
isbn = {0897912012},
publisher = {Association for Computing Machinery},
address = {New York, NY, USA},
url = {https://doi.org/10.1145/18172.18199},
doi = {10.1145/18172.18199},
booktitle = {Proceedings of the ACM SIGCOMM Conference on Communications Architectures \& Protocols},
pages = {236–240},
numpages = {5},
location = {Stowe, Vermont, USA},
series = {SIGCOMM '86}
}

@inproceedings {279972,
author = {Yi Chen and Di Tang and Yepeng Yao and Mingming Zha and XiaoFeng Wang and Xiaozhong Liu and Haixu Tang and Dongfang Zhao},
title = {Seeing the Forest for the Trees: Understanding Security Hazards in the {3GPP} Ecosystem through Intelligent Analysis on Change Requests},
booktitle = {31st USENIX Security Symposium (USENIX Security 22)},
year = {2022},
isbn = {978-1-939133-31-1},
address = {Boston, MA},
pages = {17--34},
url = {https://www.usenix.org/conference/usenixsecurity22/presentation/chen-yi},
publisher = {USENIX Association},
month = aug
}

@inproceedings{10.1145/3172871.3172879,
author = {Hayrapetian, Allenoush and Raje, Rajeev},
title = {Empirically Analyzing and Evaluating Security Features in Software Requirements},
year = {2018},
isbn = {9781450363983},
publisher = {Association for Computing Machinery},
address = {New York, NY, USA},
url = {https://doi.org/10.1145/3172871.3172879},
doi = {10.1145/3172871.3172879},
booktitle = {Proceedings of the 11th Innovations in Software Engineering Conference},
articleno = {9},
numpages = {11},
location = {Hyderabad, India},
series = {ISEC '18}
}

@inproceedings {287366,
author = {Cas Cremers and Alexander Dax and Aurora Naska},
title = {Formal Analysis of {SPDM}: Security Protocol and Data Model version 1.2},
booktitle = {32nd USENIX Security Symposium (USENIX Security 23)},
year = {2023},
isbn = {978-1-939133-37-3},
address = {Anaheim, CA},
pages = {6611--6628},
url = {https://www.usenix.org/conference/usenixsecurity23/presentation/cremers-spdm},
publisher = {USENIX Association},
month = aug
}

@inproceedings {287101,
author = {Min Shi and Jing Chen and Kun He and Haoran Zhao and Meng Jia and Ruiying Du},
title = {Formal Analysis and Patching of {BLE-SC} Pairing},
booktitle = {32nd USENIX Security Symposium (USENIX Security 23)},
year = {2023},
isbn = {978-1-939133-37-3},
address = {Anaheim, CA},
pages = {37--52},
url = {https://www.usenix.org/conference/usenixsecurity23/presentation/shi-min},
publisher = {USENIX Association},
month = aug
}

@inproceedings{10.5555/3766078.3766441,
author = {Diemunsch, Vincent and Hirschi, Lucca and Kremer, Steve},
title = {A comprehensive formal security analysis of OPC UA},
year = {2025},
isbn = {978-1-939133-52-6},
publisher = {USENIX Association},
address = {USA},
booktitle = {Proceedings of the 34th USENIX Conference on Security Symposium},
articleno = {363},
numpages = {20},
location = {Seattle, WA, USA},
series = {SEC '25}
}

@inproceedings{10.5555/3766078.3766336,
author = {Linker, Felix and Sasse, Ralf and Basin, David},
title = {A formal analysis of apple's iMessage PQ3 protocol},
year = {2025},
isbn = {978-1-939133-52-6},
publisher = {USENIX Association},
address = {USA},
booktitle = {Proceedings of the 34th USENIX Conference on Security Symposium},
articleno = {258},
numpages = {20},
location = {Seattle, WA, USA},
series = {SEC '25}
}

@inproceedings {255320,
author = {Cas Cremers and Benjamin Kiesl and Niklas Medinger},
title = {A Formal Analysis of {IEEE} 802.11{\textquoteright}s {WPA2}: Countering the Kracks Caused by Cracking the Counters},
booktitle = {29th USENIX Security Symposium (USENIX Security 20)},
year = {2020},
isbn = {978-1-939133-17-5},
pages = {1--17},
url = {https://www.usenix.org/conference/usenixsecurity20/presentation/cremers},
publisher = {USENIX Association},
month = aug
}

@inproceedings {285435,
author = {Charlie Jacomme and Elise Klein and Steve Kremer and Ma{\"\i}wenn Racouchot},
title = {A comprehensive, formal and automated analysis of the {EDHOC} protocol},
booktitle = {32nd USENIX Security Symposium (USENIX Security 23)},
year = {2023},
isbn = {978-1-939133-37-3},
address = {Anaheim, CA},
pages = {5881--5898},
url = {https://www.usenix.org/conference/usenixsecurity23/presentation/jacomme},
publisher = {USENIX Association},
month = aug
}

@misc{oscal,
  author       = {{National Institute of Standards and Technology}},
  title        = {Open Security Controls Assessment Language ({OSCAL})},
  howpublished = {\url{https://pages.nist.gov/OSCAL/}},
  year         = {2024},
  note         = {Accessed April 2026}
}

@article{li2025glue,
  title={From glue-code to protocols: A critical analysis of a2a and mcp integration for scalable agent systems},
  author={Li, Qiaomu and Xie, Ying},
  journal={arXiv preprint arXiv:2505.03864},
  year={2025}
}

@misc{saboo2026devguide,
  author = {Shubham Saboo and Kristopher Overholt},
  title = {A Developers' Guide to AI Agent Protocols},
  howpublished = {\url{https://developers.googleblog.com/developers-guide-to-ai-agent-protocols/}},
  year = {2026},
  note = {Accessed April 2026}
}

@misc{snyk2025mcp,
  author       = {Raul Onitza-Klugman},
  title        = {Prompt Injection Meets {MCP}: A New Exploitation Vector Emerging?},
  howpublished = {Snyk Labs, \url{https://labs.snyk.io/resources/prompt-injection-mcp/}},
  year         = {2025},
  month        = jul,
  note         = {Accessed June 2026}
}

@inproceedings{xia2026tracefix,
  title={TraceFix: Repairing Agent Coordination Protocols with TLA+ Counterexamples},
  author={Xia, Shuren and Li, Qiwei and Ehsan, Taqiya and Ortiz, Jorge},
  booktitle={Proceedings of the ACM Conference on AI and Agentic Systems},
  pages={181--196},
  year={2026}
}

@inproceedings{lee2025veriplan,
  title={Veriplan: Integrating formal verification and llms into end-user planning},
  author={Lee, Christine P and Porfirio, David and Wang, Xinyu Jessica and Zhao, Kevin Chenkai and Mutlu, Bilge},
  booktitle={Proceedings of the 2025 CHI Conference on Human Factors in Computing Systems},
  pages={1--19},
  year={2025}
}

@article{li2025swe,
  title={Swe-debate: Competitive multi-agent debate for software issue resolution},
  author={Li, Han and Shi, Yuling and Lin, Shaoxin and Gu, Xiaodong and Lian, Heng and Wang, Xin and Jia, Yantao and Huang, Tao and Wang, Qianxiang},
  journal={arXiv preprint arXiv:2507.23348},
  year={2025}
}

@article{gao2025more,
  title={More with less: An empirical study of turn-control strategies for efficient coding agents},
  author={Gao, Pengfei and Peng, Chao},
  journal={arXiv preprint arXiv:2510.16786},
  year={2025}
}

@misc{csa2026mcpdesignrce,
  author       = {{Cloud Security Alliance AI Safety Initiative}},
  title        = {{MCP} by Design: {RCE} Across the {AI} Agent Ecosystem},
  howpublished = {Cloud Security Alliance Lab Space, \url{https://labs.cloudsecurityalliance.org/research/csa-research-note-mcp-by-design-rce-ox-security-20260420-csa/}},
  year         = {2026},
  month        = apr,
  note         = {Accessed June 2026}
}

@misc{ox2026mcparchitecturalflaw,
  author       = {Moshe Siman Tov Bustan and Mustafa Naamnih and Nir Zadok and Roni Bar},
  title        = {The Mother of All {AI} Supply Chains: Critical, Systemic Vulnerability at the Core of {Anthropic}'s {MCP}},
  howpublished = {{OX Security} Blog, \url{https://www.ox.security/blog/the-mother-of-all-ai-supply-chains-critical-systemic-vulnerability-at-the-core-of-the-mcp/}},
  year         = {2026},
  month        = apr,
  note         = {Accessed June 2026}
}

@misc{openai2026safetybugbounty,
  author       = {{OpenAI}},
  title        = {Introducing the {OpenAI} Safety Bug Bounty Program},
  howpublished = {\url{https://openai.com/index/safety-bug-bounty/}},
  year         = {2026},
  month        = mar,
  note         = {Accessed June 2026}
}

\end{document}